\documentclass[]{emulateapj}

\usepackage{epsfig}

\usepackage{amssymb}

\usepackage{multirow}

\usepackage{enumitem}

\usepackage{ulem}

\usepackage{rotating}
\usepackage{natbib}
\usepackage{latexsym}
\bibpunct{(}{)}{;}{a}{}{,}
\usepackage{url}

\begin{document}

\newcommand{\ledd}{%
$L_\mathrm{Edd}$}

\newcommand{\IGR}{IGR~J18245--2452}

\newcommand{\Msun}{M$_\mathrm{\odot}$}

\def\rem#1{{\bf #1}}
\def\hide#1{}

\def \aj {AJ}
\def \mnras {MNRAS}
\def \apj {ApJ}
\def \apjs {ApJS}
\def \apjl {ApJL}
\def \aap {A\&A}
\def \aapr {A\&ARv}
\def \nat {Nature}
\def \araa {ARAA}
\def \pasp {PASP}
\def \aaps {AAPS}
\def \prd {PhRvD}
\def \apss {ApSS}

\newcommand{\specialcell}[2][c]{%
  \begin{tabular}[#1]{@{}c@{}}#2\end{tabular}}

\title{X-ray states of redback millisecond pulsars}

\author{M. Linares\altaffilmark{1,2}}
%


\submitted{Accepted for publication in The Astrophysical Journal}

\altaffiltext{1}{Instituto de Astrof{\'i}sica de Canarias, c/ V{\'i}a
  L{\'a}ctea s/n, E-38205 La Laguna, Tenerife, Spain}
\altaffiltext{2}{Universidad de La Laguna, Departamento de
  Astrof{\'i}sica, E-38206 La Laguna, Tenerife, Spain}
%
%



\begin{abstract}

Compact binary millisecond pulsars with main-sequence donors, often
referred to as ``redbacks'', constitute the long-sought link between
low-mass X-ray binaries and millisecond radio pulsars, and offer a
unique probe of the interaction between pulsar winds and accretion
flows.
We present a systematic study of eight nearby redbacks, using more
than 100 observations obtained with {\it Swift}'s X-ray Telescope.
We distinguish between three main states: pulsar, disk and outburst
states.
We find X-ray mode switching in the disk state of PSR~J1023+0038 and
XSS~J12270--4859, similar to what was found in the other redback which
showed evidence for accretion: rapid, recurrent changes in X-ray
luminosity (0.5--10~keV, L$_\mathrm{X}$), between
[6--9]$\times$10$^{32}$~erg~s$^{-1}$ (disk-passive state) and
[3--5]$\times$10$^{33}$~erg~s$^{-1}$ (disk-active state).
This strongly suggests that mode switching---which has not been
observed in quiescent low-mass X-ray binaries---is universal among
redback millisecond pulsars in the disk state.
We briefly explore the implications for accretion disk truncation, and
find that the inferred magnetospheric radius in the disk state of
PSR~J1023+0038 and XSS~J12270--4859 lies outside the light cylinder.
Finally, we note that all three redbacks which have developed
accretion disks have relatively high L$_\mathrm{X}$ in the pulsar
state ($>$10$^{32}$~erg~s$^{-1}$).

\end{abstract}

\maketitle

\section{Introduction}
\label{sec:intro}

During the last five years, the connection between rotation-powered
millisecond radio pulsars (MRPs) and accretion-powered low-mass X-ray
binaries (LMXBs) has been firmly established on an observational
basis.
Conclusive evidence for accretion has been found in three systems that
have also been directly detected as rotation-powered MRPs:
PSR~J1023+0038 \citep[J1023 hereafter;][]{Archibald09},
PSR~J1824--2452I/IGR~J18245--2452 \citep[M28-I hereafter, in the
  globular cluster M28;][]{Papitto13b} and XSS~J12270--4859
\citep[J12270 hereafter;][]{deMartino10,deMartino13,Roy14}.
This is the strongest confirmation of the so-called ``recycling
scenario'' for MRP formation, where neutron stars are spun up by the
accretion of matter in an LMXB phase \citep{Alpar82}. It has also
become clear \citep{Papitto13b,Stappers14} that the evolution from the
LMXB to the MRP phase does not involve a sharp ``accretion turn-off'',
and multiple MRP$\leftrightarrow$LMXB transitions can actually take
place \citep[as predicted by some evolutionary models,][]{Tauris12}.

In all three LMXB-MRP transition pulsars (J1023, M28-I and J12270) the
neutron star is in a close orbit (period
P$_\mathrm{orb}$$\simeq$4--11~hr) with a low-mass main-sequence
companion star (with mass
M$_\mathrm{C}$$\gtrsim$0.1--0.5~M$_\mathrm{\odot}$).
From the radio pulsar point of view, this classifies all three
transition pulsars as ``redbacks'' \citep{Roberts13}, a name inspired
by the strong irradiation of the companion in analogy with the
``black widow'' pulsars \citep[which have $\sim$10 times lower
  M$_\mathrm{C}$;][]{Fruchter88}.
All three transition pulsars are redbacks, but not all known redbacks
have been observed in an accretion state.
It is therefore natural to study in depth all redback MRPs
to understand how the LMXB-MRP transition occurs and to search for
more transition systems.

Non-thermal X-ray emission modulated at the orbital period has been
observed in both black widow \citep{Stappers03} and redback MRPs
\citep[in the ``pulsar state'', see Secs.~\ref{sec:results} \&
  \ref{sec:conclusions} for
  definition;][]{Bogdanov05,Bogdanov11b,Bogdanov14}, with X-ray
luminosities (L$_\mathrm{X}$, 0.5--10~keV) of
$\sim$10$^{31}$--10$^{32}$~erg~s$^{-1}$.
This is typically ascribed to Doppler-boosted and partially occulted
intrabinary shock emission \citep{Arons93}.
Thermal pulsed X-rays from the heated polar caps are also detected
from some MRPs, with typical
L$_\mathrm{X}$$\sim$10$^{30}$~erg~s$^{-1}$.
Similarly, the quiescent X-ray spectrum of neutron star LMXBs can be
broadly divided into a thermal component powered by incandescent
crustal emission or residual surface accretion
\citep{Brown98,Wijnands01}, and a non-thermal component of less
certain origin \citep{Campana98}.

Using {\it Chandra} observations of M28-I in quiescence,
\citet{Linares14} discovered X-ray mode switching: rapid transitions
between active (L$_\mathrm{X}$$\sim$10$^{33}$~erg~s$^{-1}$) and
passive (L$_\mathrm{X}$$\sim$10$^{32}$~erg~s$^{-1}$) states.
They interpreted these as fast transitions between magnetospheric
accretion and pulsar wind shock emission as the neutron star
magnetospheric radius moves in and out of the light cylinder (see
\citealt{Linares14}, for details; see also \citealt{Ferrigno14}, for
peculiar variability at higher L$_\mathrm{X}$).
Besides M28-I \citep{Begin06}, a number of redbacks have been found in
globular clusters (GCs) \citep{Camilo00,DAmico01,Possenti03}. Even if
they may also occasionally undergo accretion episodes,
multi-wavelength monitoring of MRPs in GCs is challenging due to a
combination of large distance, strong absorption, crowded fields and
acceleration within the cluster.

Thanks to the numerous GeV sources discovered with {\it Fermi}'s Large
Area Telescope \citep[LAT,][]{Abdo10,Nolan12}, a new population of
nearby redback MRPs in the Galactic plane has been uncovered
\citep{Hessels11,Ray12,Roberts13}.
Like J1023 and J12270, these recently discovered redbacks have the
advantage of being nearby (distance D$\lesssim$4~kpc), barely absorbed
(equivalent hydrogen column densities
N$_\mathrm{H}$$\lesssim$10$^{21}$~cm$^{-2}$) and in non-crowded
fields.
This makes them accesible to a broader range of space and ground-based
observatories, allowing much more frequent monitoring.

In this work we assemble the current sample of nearby redbacks in the
Galactic field, a total of eight systems including J1023 and J12270,
and study their properties using all available observations taken with
the X-ray telescope (XRT) aboard the {\it Swift} observatory.
We define three main states of redback MRPs: pulsar, disk and outburst
state\footnote{Note that the presence of an accretion disk in the disk
  state has been established by the detection of broad double-peaked
  optical emission lines, even if an accretion disk component has not
  been detected to date in its X-ray spectrum. See, e.g.,
  \citet{Archibald09} and Sections~\ref{sec:accretors} \&
  \ref{sec:disk}.}.
We find X-ray mode switching in the disk state of J1023 and J12270
\citep{Patruno14,deMartino13}, in clear analogy with M28-I
\citep{Linares14}.
We also study the luminosity of the pulsar state using the full sample
of redbacks, report two candidates for the X-ray counterpart of
PSR~J2129--0429 and suggest two promising targets to search for future
transitions to the disk state: PSR~J1723--2837 and PSR~J2215+5135.
Our main results and conclusions are summarized in
Section~\ref{sec:conclusions}.

\section{Analysis}
\label{sec:data}

We analyzed all {\it Swift}-XRT observations of the following eight
redbacks available in March 2014: J1023, J12270, PSR~J1628--32 (J1628
hereafter), PSR~J1723--28 (J1723), PSR~J1816+4510 (J1816),
PSR~J2129--0429 (J2129), PSR~J2215+5135 (J2215) and PSR~J2339--0533
(J2339; see details and references for each source in
Table~\ref{table:redbacks} \& Section~\ref{sec:results}).
This yielded 107 observations with a total on-source exposure time of
190~ksec.
All observations were taken in imaging mode (photon counting, PC)
except for nine short non-imaging pointings that were not included in
this analysis (windowed timing, totalling less than 0.9~ksec).
Table~\ref{table:redbacks} shows a summary of the resulting source
sample and dataset as well as the values of D and N$_\mathrm{H}$ used
in this work.

All data and auxilliary products were created using generic (v. 6.12)
and {\it Swift}-specific (v. 3.9.28) \textsc{Ftools} with the latest
CALDB calibration files, after creating reprocessed event files using
\textsc{xrtpipeline} (v. 0.12.6).
In the case of J2129 (Sec.~\ref{sec:j2129}), as no accurate
(arcsecond) location has been reported, we calculated positions for
our two counterpart candidates using {\it Swift} UKSSDC's online
tool\footnote{\url{http://www.swift.ac.uk/user\_objects/}}, with both
the point-spread function (PSF) and the UVOT-enhanced methods
\citep[][which corrects the absolute astrometry using nearby UV
  sources with accurate locations]{Goad07,Evans09}.
We also verified that the positions given by \textsc{xrtcentroid}
(v. 0.2.9) are consistent with the UVOT-enhanced locations
reported below.

We extracted source and background spectra within \textsc{Xselect}
(v. 4.2b) using circular regions with 20--30'' and 60'' radius,
respectively.
We created ancilliary response files using exposure maps (combining
multiple observations within \textsc{Ximage}, v. 4.5.1, when
necessary), thereby taking into account dead pixels/columns and XRT's
PSF.
Using the latest response matrices (v. 014) from the \textsc{caldb}
database, we fitted all 0.5--10~keV source spectra within
\textsc{Xspec} \citep[v. 12.7.1,][]{Arnaud96} grouping the spectra to
a minimum of 10 counts per spectral bin, or using Cash's C-statistic
for spectra with low number of source counts ($\lesssim$50).

Errors on spectral parameters are given at the 68\% confidence level
(c.l.).
When N$_\mathrm{H}$ was unconstrained in the XRT spectral fits, we
fixed it to literature values if available, or to the total Galactic
value in the line of sight \citep[][see
  Table~\ref{table:redbacks}]{Kalberla05}.
For the two sources not detected, we calculated 90\% c.l.
L$_\mathrm{X}$ upper limits using a 30'' region, the D and
N$_\mathrm{H}$ values given in Table~\ref{table:redbacks} and assuming
a power law spectrum with photon index $\Gamma$ in the 1--2 range.
In only three cases (J1628, J2129 and J2215) we use D measurements
that rely exclusively on the pulsar's dispersion measure
(\citealt{Ray12}, which at high Galactic latitudes may underestimate D
by a factor 2, \citealt{Roberts11}).
The distance to J12270 is only constrained between 1.4 and 3.6~kpc
\citep{deMartino13}. For this particular source we use the full
1.4--3.6~kpc range when quoting L$_\mathrm{X}$ uncertainties, although
we favor the shortest D value as explained in
Section~\ref{sec:accretors}.
In all other sources we use the statiscal uncertainty on the flux
measurements when quoting L$_\mathrm{X}$ uncertainties.

\begin{figure*}
  \begin{center}
  \resizebox{1.0\columnwidth}{!}{\rotatebox{-90}{\includegraphics[]{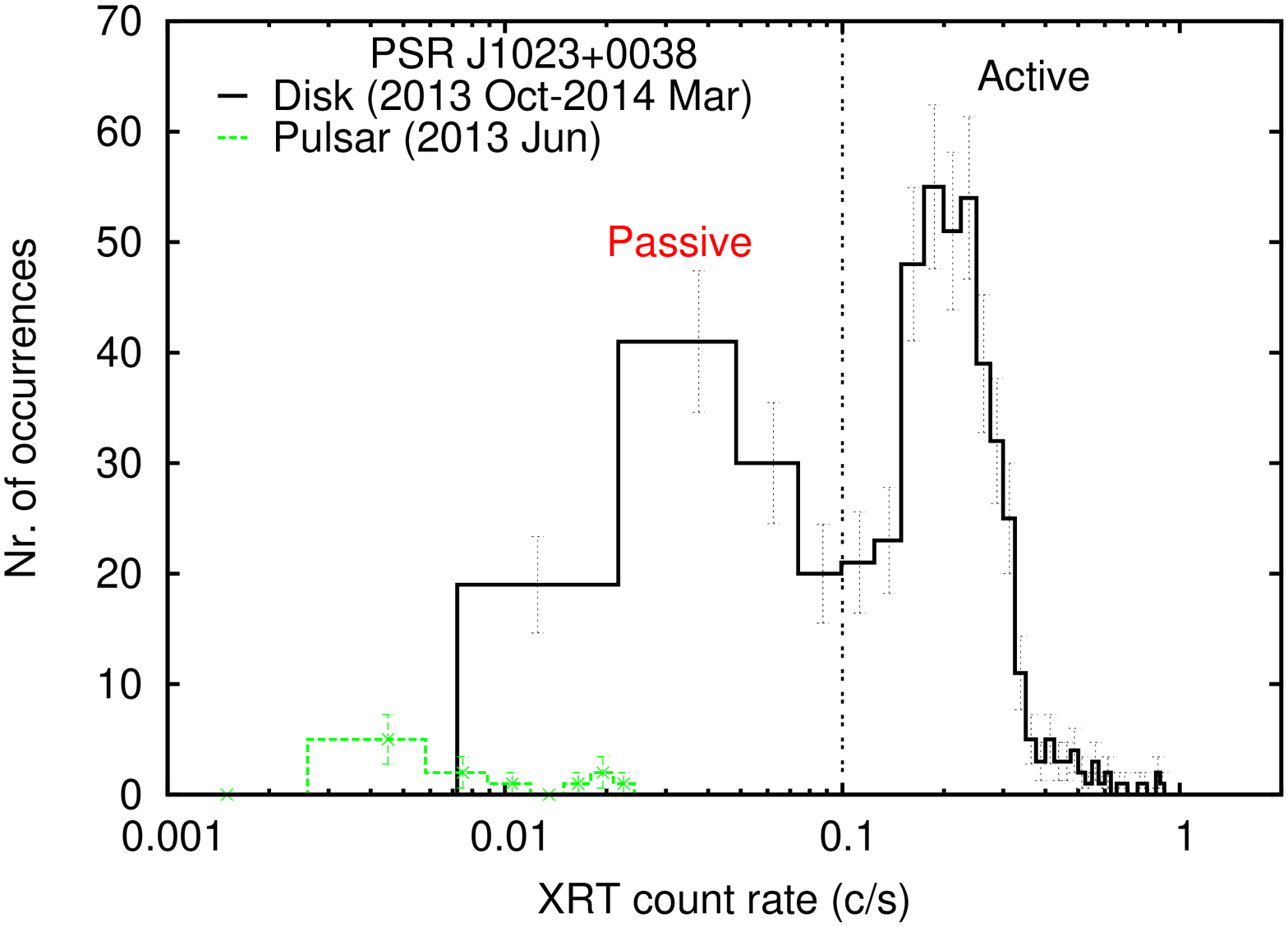}}}
  \resizebox{1.0\columnwidth}{!}{\rotatebox{-90}{\includegraphics[]{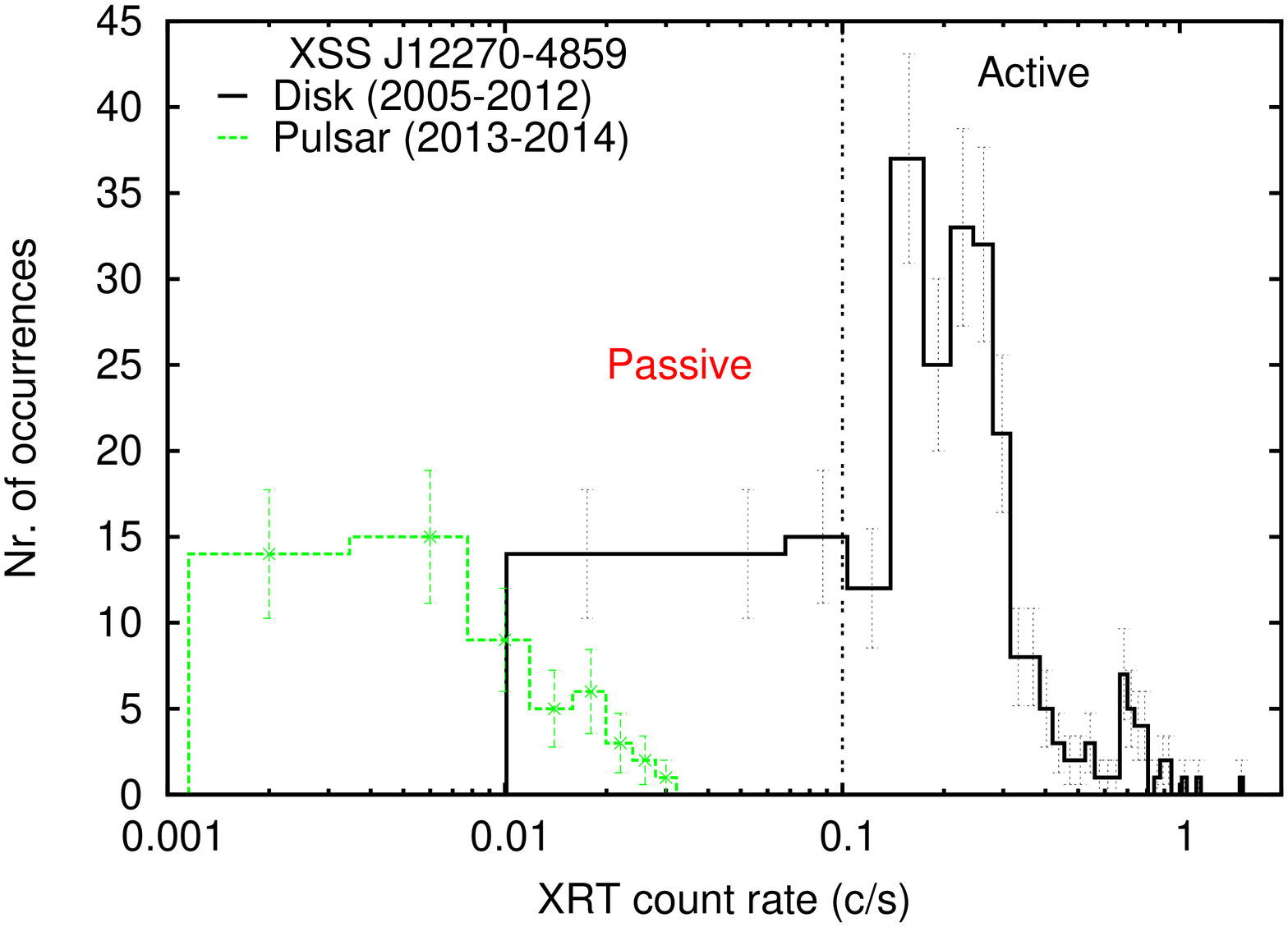}}}
  \caption{
Count rate distribution in the 0.3--10~keV XRT light curves of J1023
{\it (left)} and J12270 {\it (right)}.
The disk state (solid black histogram) and pulsar state (dashed green
histogram) are shown separately.
The horizontal dotted line at 0.1~c~s$^{-1}$ marks the boundary
between disk-active and disk-passive states (labeled in black and red,
respectively).
} %
    \label{fig:hist}
 \end{center}
\end{figure*}

\begin{figure*}
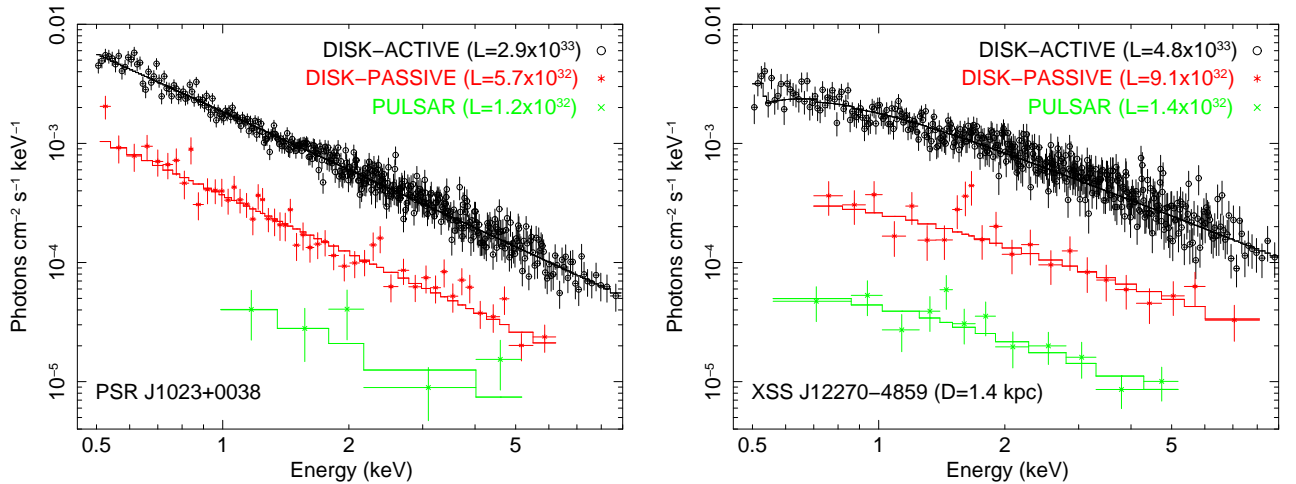

  \begin{center}
  \resizebox{1.0\columnwidth}{!}{\rotatebox{-90}{\includegraphics[]{f2a.eps}}}
  \resizebox{1.0\columnwidth}{!}{\rotatebox{-90}{\includegraphics[]{f2b.eps}}}
  \caption{
Unfolded XRT spectra of J1023 {\it (left)} and J12270 {\it (right)},
together with the best fit powerlaw models, for the disk-active (black
circle), disk-passive (red asterisk) and pulsar (green x) states.
The 0.5--10~keV luminosity of each state is indicated, in erg~s$^{-1}$
(see text for details).
The faintest spectra are binned for display purpose only
  (Sec.~\ref{sec:data}, Table~\ref{table:redbacks}).
} %
    \label{fig:spec}
 \end{center}
\end{figure*}

We also extracted barycentered light curves in the full (0.3--10~keV)
XRT energy band from the two brightest sources, J1023 and J12270,
using time bins in the 150--300~s range and a 30'' radius extraction
region.
Dead pixels and columns combined with different pointings can
introduce XRT count rate variations that are not intrinsic to the
source. We verified that such exposure correction (which we apply to
the energy spectra as explained above) does not change the count rate
distribution in the light curves significantly.
To do so, we corrected each orbit's light curve separately applying
correction factors derived from the corresponding exposure maps (using
\textsc{xrtlccorr} and
\textsc{lcmath})\footnote{\url{http://www.swift.ac.uk/analysis/xrt/lccorr.php}}.
This correction typically increases the count rate by 20--30\% (about
10 times more in a few orbits with bad geometry, where several dead
columns cross the source extraction region).
We find that the count rate distributions obtained from the
exposure-corrected and original/uncorrected light curves are fully
consistent, and we use the latter herafter.
The light curves are not background subtracted; we measure an
average background rate using a 30'' region of 7$\times$10$^{-4}$ and
5$\times$10$^{-4}$~c/s in J1023 and J12270, respectively, i.e. more
than ten times lower than the collected source count rates.

The two brightest redbacks, J1023 and J12270, have been observed with
{\it Swift} in both the {\it pulsar state} (when radio pulsations are
detected, despite being eclipsed during a fraction of the orbit) and
in the {\it disk state} \citep[when L$_\mathrm{X}$ is $\sim$10$\times$
  higher and double-peaked broad emission lines are observed;][see
  Table~\ref{table:redbacks} for dates and full
  references]{Archibald09,Patruno14,deMartino10,Bassa14}.
We treated these subsets of observations independently, accumulating
spectra and light curves for the pulsar and disk states separately.
Note that the {\it Swift} observations were taken over the course of
five years under different proposals with heterogeneous sampling and
observing strategy. As a result, the accumulated exposure of disk and
pulsar states in J1023 and J12270 (Figure~\ref{fig:hist},
Table~\ref{table:redbacks}) is not representative of the intrinsic
duty cycle of these states.
The disk state of J1023 and J12270, however, has been extensively
observed by {\it Swift} (48 and 18 observations, respectively). Thus
we stress that the double-peaked count rate histograms that we find
and report in Section~\ref{sec:accretors} are not due to a sampling
effect.


\section{Results and Discussion}
\label{sec:results}

The results of our global {\it Swift} study allow a clear distinction
between three main states of redback MRPs, in order of increasing
L$_\mathrm{X}$: i) pulsar state; ii) disk state and iii) outburst
state.
These results are summarized in Table~\ref{table:redbacks}, which
gives also references to previous work.
We present in Section~\ref{sec:accretors} our results on the only two
nearby redbacks that have shown accretion episodes, J1023 and J12270,
which reveal remarkably similar X-ray properties.
In the remaining 6 redbacks, discussed in
Sections~\ref{sec:j1628}--\ref{sec:j2339}, there are no reports of
disk emission lines, complete disappearance of radio pulsations,
enhanced X-ray luminosity, X-ray bursts, pulsations or any other
accretion-powered phenomenon.
This strongly suggests that, to date, they have always been observed
in the pulsar state.
Sections~\ref{sec:disk} and \ref{sec:pulsar} extend our discussion of
the disk and pulsar states, respectively.

\begin{table*}
\center
\footnotesize
\caption{Luminosities and spectra of eight nearby redback pulsars observed with {\it Swift}-XRT.}
\begin{minipage}{\textwidth}
\begin{tabular}{l c c c c c c c c}
\hline\hline
\footnotetext{N$_\mathrm{obs}$: number of {\it Swift} observations
  available in March 2014, analyzed in this work ($^*$same set of
  observations where J1023 and J12270 switch between disk-active and
  disk-passive states). Exp.: Accumulated {\it Swift}-XRT (PC mode)
  exposure for each source/state. Rate: Net 0.5--10~keV XRT (PC mode)
  count rate. L$_\mathrm{X}$: 0.5--10~keV luminosity using the
  distances given below (D). $\Gamma$: Power law index (photon flux
  $\propto$ E$^{-\Gamma}$). N$_\mathrm{H}$: absorbing column density,
  fixed at the literature values given below when unconstrained
  ($\equiv$). OX: other references to previous X-ray studies. Errors
  ($\pm$) and upper limits ($<$) are given at the 68\% and 90\% c.l.,
  respectively. Spectra with low number of counts ($\lesssim$50) were
  fitted using {\it Xspec}'s C-statistic (marked $^{CS}$ in the
  Table).}
State & N$_\mathrm{obs}$ & Dates & Exp. & Rate & L$_\mathrm{X}$ & $\Gamma$ & N$_\mathrm{H}$ & $\chi^2$/dof \\
  & & & (ksec) & (c~s$^{-1}$) & (0.5--10~keV; erg~s$^{-1}$)  & & (10$^{21}$ cm$^{-2}$) & \\
\hline
\multicolumn{9}{c}{\textbf{PSR~J1023+0038} \footnote{(D=1.37~kpc, \citealt{Deller12}; N$_\mathrm{H}$$<$1.5$\times$10$^{19}$~cm$^{-2}$, \citealt{Bogdanov11b}; OX: \citealt{Patruno14}, \citealt{Archibald10}, \citealt{Stappers14}, \citealt{Takata14})}}\\
\hline
Disk-Active & 43$^*$ & 2013 Oct--2014 Mar & 64.9 &  2.2$\times$10$^{-1}$ & [2.85$\pm$0.04]$\times 10^{33}$  & 1.62$\pm$0.02 & $<$0.09 & 376.4/378 \\
Disk-Passive & 43$^*$ & 2013 Oct--2014 Mar & 18.7 &  4.4$\times$10$^{-2}$ & [5.7$\pm$0.3]$\times 10^{32}$ & 1.62$\pm$0.07 & $\equiv$0.015 & 43.6/48 \\
Pulsar & 2 & 2013 Jun & 3.8 & 7.6$\times$10$^{-3}$ & [1.6$\pm$0.5]$\times 10^{32}$ & 0.9$\pm$0.4 & $\equiv$0.015 & 25.9/27$^{CS}$ \\
\hline
\multicolumn{9}{c}{\textbf{XSS~J12270--4859} \footnote{(D=1.4--3.6~kpc, \citealt{deMartino13}; N$_\mathrm{H}$=6.1$\times$10$^{20}$~cm$^{-2}$, \citealt{Bogdanov14}; OX: \citealt{deMartino10}, \citealt{Bassa14})}}\\
\hline
Disk-Active & 18$^*$ & 2005-2012 & 26.0 &  3.0$\times$10$^{-1}$ & [4.8--32]$\times 10^{33}$  & 1.36$\pm$0.03 & 0.8$\pm$0.1 & 395.5/393 \\
Disk-Passive & 18$^*$ & 2005-2012 & 4.8 &  5.0$\times$10$^{-2}$ & [9.1--60]$\times 10^{32}$ & 1.15$\pm$0.12 & $\equiv$0.8 & 17.6/21 \\
Pulsar & 12 & 2013-2014 & 17.5 & 7.5$\times$10$^{-3}$ & [1.5--9.8]$\times 10^{32}$ & 1.2$\pm$0.2 & $\equiv$0.8 & 6.6/10 \\
\hline
\multicolumn{9}{c}{\textbf{PSR~J1628--32} \footnote{(D=1.2~kpc, \citealt{Ray12}; N$_\mathrm{H}$=1.5$\times$10$^{21}$~cm$^{-2}$, \citealt{Kalberla05})}}\\
\hline
Pulsar & 8 & 2013 Apr-Jun & 3.2 & $<$1.7$\times$10$^{-3}$ & $<$2.2$\times 10^{31}$ & -- & $\equiv$1.5 & -- \\
\hline
\multicolumn{9}{c}{\textbf{PSR~J1723--28} \footnote{(D=0.75~kpc, \citealt{Crawford13}; N$_\mathrm{H}$=3.4$\times$10$^{21}$~cm$^{-2}$, \citealt{Bogdanov14b}; OX: \citealt{Hui14})}}\\
\hline
Pulsar & 2 & 2010 Mar & 6.6 & 2.0$\times$10$^{-2}$ & [2.4$\pm$0.6]$\times 10^{32}$ & 0.9$\pm$0.2 & $\equiv$3.4 & 10.0/10 \\
\hline
\multicolumn{9}{c}{\textbf{PSR~J1816+4510} \footnote{(D=4.5~kpc, \citealt{Kaplan13}; N$_\mathrm{H}$=0.3$\times$10$^{21}$~cm$^{-2}$, \citealt{Kalberla05}; OX: \citealt{Kaplan12})}}\\
\hline
Pulsar & 4 & 2010 Aug & 4.7 & $<$8.2$\times$10$^{-4}$ &  $<$1.3$\times 10^{32}$ & -- & $\equiv$0.3 & -- \\
\hline
\multicolumn{9}{c}{\textbf{PSR~J2129--0429 A} \footnote{(D=0.9~kpc, \citealt{Ray12}; N$_\mathrm{H}$=0.3$\times$10$^{21}$~cm$^{-2}$, \citealt{Kalberla05}. Both X-ray counterpart candidates shown, see Sec.~\ref{sec:j2129})}}\\
\hline
Pulsar & 1 & 2010 Dec & 10.4 & 6.0$\times$10$^{-3}$ &  [4.2$\pm$0.6]$\times 10^{31}$ & 1.8$\pm$0.2 & $\equiv$0.3 & 37.9/54$^{CS}$ \\
\hline
\multicolumn{9}{c}{\textbf{PSR~J2129--0429 B} $^f$}\\
\hline
Pulsar & 1 & 2010 Dec & 10.4 & 2.8$\times$10$^{-3}$ & [2.6$\pm$0.8]$\times 10^{31}$ & 1.0$\pm$0.3 & $\equiv$0.3 & 30.6/27$^{CS}$ \\
\hline
\multicolumn{9}{c}{\textbf{PSR~J2215+5135} \footnote{(D=3~kpc, \citealt{Ray12}; N$_\mathrm{H}$=2.1$\times$10$^{21}$~cm$^{-2}$, \citealt{Gentile14})}}\\
\hline
Pulsar & 2 & 2010 Jul, 2013 Dec & 15.9 & 1.5$\times$10$^{-3}$ &  [1.3$\pm$0.4]$\times 10^{32}$ & 1.8$\pm$0.4 & $\equiv$2.1 & 27.3/22$^{CS}$ \\
\hline
\multicolumn{9}{c}{\textbf{PSR~J2339--0533} \footnote{(D=1.1~kpc, N$_\mathrm{H}$$<$1.6$\times$10$^{21}$~cm$^{-2}$, \citealt{Romani11}; OX: \citealt{Kong12})}}\\
\hline
Pulsar & 15 & 2009-2012 & 37.4 & 3.0$\times$10$^{-3}$ & [2.7$\pm$0.5]$\times 10^{31}$ & 1.4$\pm$0.3 & $<$1.2 & 7.5/6 \\
\hline\hline
\multicolumn{9}{c}{\textbf{cf. IGR~J18245--2452 (M28-I)} \footnote{For comparison, we list here the luminosity and spectral parameters of M28-I measured with {\it Chandra}, taken from \citet{Linares14} (D=5.5~kpc, \citealt{Harris96}, 2010 revision)}}\\
\hline
Disk-Active &  & 2008,2013 & &   & [3.9$\pm$0.1]$\times 10^{33}$  & 1.51$\pm$0.04 & 2.9$\pm$0.2 & 33.3/27 \\
Disk-Passive &  & 2008 &  & & [5.6$\pm$1.0]$\times 10^{32}$ & 1.45$\pm$0.15 & 2.6$\pm$0.8 & 29.7/29 \\
Pulsar &  & 2002 & &  & [2.2$\pm$0.4]$\times 10^{32}$ & 1.2$\pm$0.2 & $\equiv$2.6 & 294.2/647$^{CS}$ \\
\hline\hline
\end{tabular}
\end{minipage}
\label{table:redbacks}
\end{table*}


\subsection{PSR~J1023+0038 and XSS~J12270--4859}
\label{sec:accretors}

J1023 was discovered as a 1.69-ms MRP in 2007 \citep[][with
  P$_\mathrm{orb}$=0.20~d]{Archibald09}, yet it had an accretion disk
in 2000-2001 \citep{Wang09}.
J12270 was the first GeV source identified as an LMXB
\citep{deMartino10,Hill11,deMartino13} and optical observations
revealed its P$_\mathrm{orb}$=0.29~d \citep{deMartino13b}.
Their paths crossed in 2013, when J1023 reentered the disk state
\citep{Stappers14,Patruno14} and J12270 transitioned to a MRP phase,
being observed as a rotation-powered pulsar for the first time
\citep[with also P$_\mathrm{s}$=1.69~ms;][]{Roy14,Bassa14}.

Figure~\ref{fig:hist} shows the count rate histograms obtained from
the light curves of the disk and pulsar states of J1023 and J12270.
The distribution of count rates in the disk state of J1023 is bimodal,
with two clearly distinct peaks centered around 0.04 and 0.2
c~s$^{-1}$ (Fig.~\ref{fig:hist}, left).
This behavior is analogous to the mode switching observed in the
disk state of M28-I \citep[][see their Figure~6]{Linares14}.
The disk state of J12270 (with $\sim$2.7 times shorter exposure) shows
a similar count rate distribution (Fig.~\ref{fig:hist}, right): a peak
around 0.2~c~s$^{-1}$ and a broader component at
0.01--0.1~c~s$^{-1}$.

Using this bimodal count rate distribution, and following the
nomenclature of \citet{Linares14}, we divide the disk state into {\it
  disk-active} ($>$0.1~c~s$^{-1}$) and {\it disk-passive}
($<$0.1~c~s$^{-1}$) states.
We find that during the {\it Swift} observations taken in the disk
state, J1023 and J12270 were in the disk-active state for about 78\%
and 84\% of the time, respectively.
In contrast, M28-I was in the disk-active state for only $\sim$60\% of
the 2008 {\it Chandra} observations \citep{Linares14}.
Because {\it Swift}'s uninterrupted observing windows are short
(typically $\sim$1--2~ksec) compared to the duration of disk-active and
disk-passive states (tens of ksec), mode switching is best seen after
accumulating many observations.
The pulsar state observations are fainter in both sources, with
average count rates 6--7 times lower than the disk-passive state.

The {\it Swift}-XRT spectra of the disk-active, disk-passive and
pulsar states, shown in Figure~\ref{fig:spec}, are all fitted
satisfactorily with a simple absorbed power law model (reduced
chi-squared between 0.7 and 1, Table~\ref{table:redbacks}).
Despite the low N$_\mathrm{H}$, we do not find evidence of a thermal
component in the disk-active state, which has the highest
signal-to-noise ratio.
Using a neutron star atmosphere model \citep[][with mass and radius
  fixed at 1.4M$_\odot$ and 10km, respectively]{Heinke06b}, we place
upper limits on the intrinsic effective temperature of
7.3$\times$10$^5$~K and 9.1$\times$10$^5$~K for J1023 and J12270,
respectively \citep[a fainter thermal component has been reported
  by][from {\it Chandra} observations of J1023]{Bogdanov11}.
We report details of the spectral fits in Table~\ref{table:redbacks},
including all best-fit spectral parameters and L$_\mathrm{X}$.
In summary, the {\it Swift}-XRT spectra of J1023 and J12270 are
consistent with those measured with {\it Chandra} in M28-I: purely
non-thermal X-ray spectra with a power law photon index
$\Gamma$$\simeq$1--1.5 (see Figure~\ref{fig:spec} and
Table~\ref{table:redbacks}).

The luminosities of the disk-active, disk-passive and pulsar states of
J1023 and J12270 (Table~\ref{table:redbacks}) are also consistent with
the luminosities of the corresponding states in M28-I
\citep{Linares14}.
We therefore find conclusive evidence for {\it mode switching} in the
disk state of J1023: a bimodal count rate distribution with
non-thermal spectra and luminosity fully consistent with M28-I.
This confirms the earlier suggestion based on the first {\it Swift}
observation of the disk state of J1023 \citep{Patruno14}.
Our {\it Swift} results \citep[as well as the {\it XMM} observations
  presented in][]{deMartino10}, reveal that mode switching is also
present in the disk state of J12270.
Moreover, short X-ray flares are superposed on the disk-active state
light curves of J12270 \citep[see][for an extended
  discussion]{deMartino10} and J1023 (Bogdanov, priv. comm.).
These flares are most likely responsible for the high count rate tails
in the histograms shown in Figure~\ref{fig:hist}, as well as the small
peak in the count rate distribution of the disk-active state of J12270
at $\sim$0.7~c~s$^{-1}$ (Fig.~\ref{fig:hist}, right).

\begin{figure*}
  \begin{center}
  \resizebox{1.65\columnwidth}{!}{\rotatebox{0}{\includegraphics[]{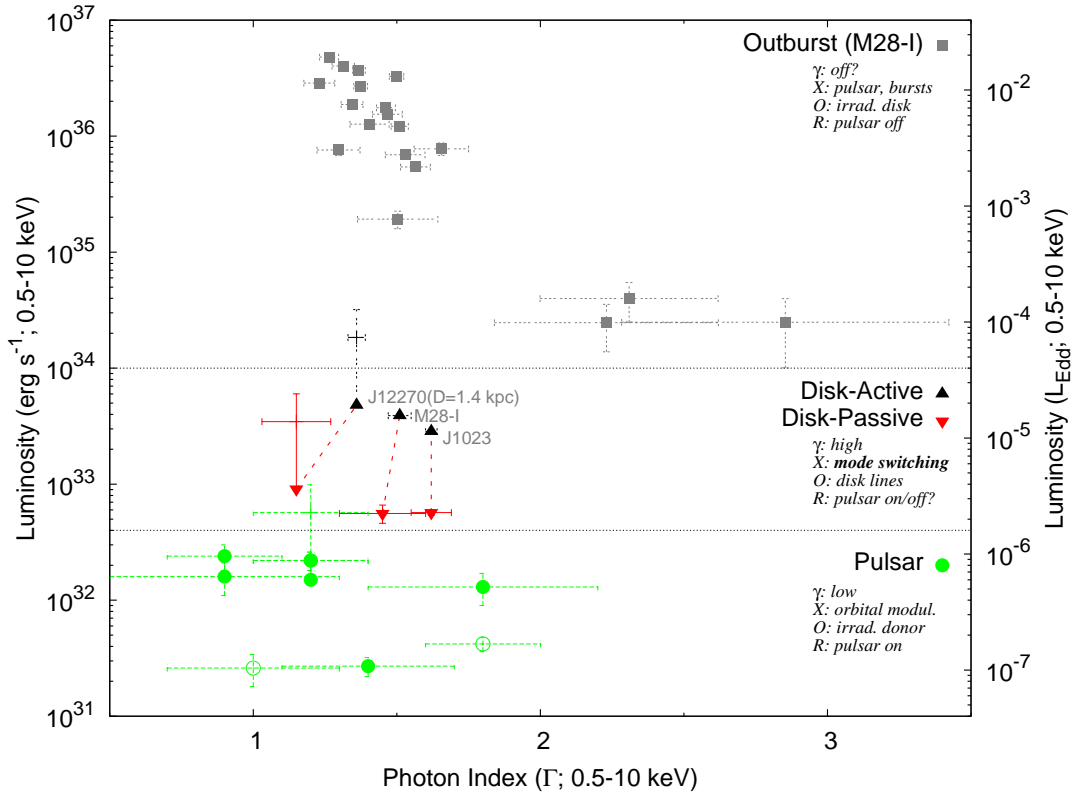}}}
  \caption{
X-ray luminosity and photon index for seven redback millisecond
pulsars in their three main states: pulsar (green circles; see
Fig.~\ref{fig:pulsar} for source identification), disk (triangles) and
outburst (gray squares).
Approximate L$_\mathrm{X}$ ranges are indicated with horizontal dotted
lines.
Mode switching is always observed in the disk state, defining
disk-active (black upward triangles) and disk-passive (red downward
triangles) states, connected with red dashed lines.
The main features of each state are outlined along the right axis, in
the $\gamma$-ray ($\gamma$), X-ray (X), optical (O) and radio (R)
bands.
} %
    \label{fig:lg}
 \end{center}
\end{figure*}

Figure~\ref{fig:lg} presents L$_\mathrm{X}$ and $\Gamma$ for all
redbacks detected with {\it Swift} in the pulsar and disk states. For
comparison, we also show the {\it Chandra} results on M28-I, including
the outburst state.
Mode switching has now been observed in all three MRPs that have shown
accretion episodes: M28-I, J1023 and J12270.
Therefore, based on the currently available data, we conclude that
mode switching is an ubiquitous phenomenon in the disk state of
redback MRP-LMXB transition systems (see Sec.~\ref{sec:disk} for
further discussion).
In Figure~\ref{fig:lg} we also show the approximate boundaries between
outburst and disk states (L$_\mathrm{X}$$\sim$10$^{34}$~erg~s$^{-1}$)
and between disk and pulsar states
(L$_\mathrm{X}$$\sim$4$\times$10$^{32}$~erg~s$^{-1}$), as well as a
summary of the multi-wavelength properties of the three states.

It is interesting to note that the gamma-ray luminosity exceeds
  L$_\mathrm{X}$ in the pulsar, disk-passive and disk-active states
  \citep[by a factor $\sim$7, $\sim$10 and $\sim$2, respectively;
    using $\sim$0.1--300~GeV luminosities
    from][]{Tam10,Hill11,Stappers14}.
See the works by \citet{Papitto14} and \citet{Takata14} for possible
mechanisms to explain such gamma-ray emission, brighter in the disk
state than in the pulsar state, but so far undetected in the
outburst state.

If the distance to J12270 is 1.4~kpc (at the low end of the range
estimated by \citealt{deMartino13}) the luminosities of the
disk-active, disk-passive and pulsar states of J12270 agree with those
found in M28-I and J1023 (whose distances are well known from studies
of the globular cluster and from parallax measurements, respectively).
This can be seen in Table~\ref{table:redbacks} and
Figure~\ref{fig:lg}.
Our results thus favor the 1.4~kpc distance estimate for J12270. For
this reason we plot J12270's L$_\mathrm{X}$ using D=1.4~kpc (but we
show with error bars the full L$_\mathrm{X}$ range allowed by the
uncertainty in D).
In all three systems the disk-passive state is 5--7 times fainter
(lower L$_\mathrm{X}$) than the disk-active state, but 3--6 times
brighter (higher L$_\mathrm{X}$) than the pulsar state.

\subsection{PSR~J1628--3205}
\label{sec:j1628}

Little is known about J1628 apart from its basic pulsar parameters
\citep[P$_\mathrm{orb}$$\simeq$0.21~d and
  P$_\mathrm{s}$$\simeq$3.21~ms,][]{Ray12} and its association with
the {\it Fermi}-LAT source 2FGL~J1628.3-3206 (1FGL~J1627.8-3204).
No X-ray or optical counterpart has been published at the time of
writing, hence only the LAT (arcminute) location is available.

The eight {\it Swift}-XRT observations cover alltogether $\sim$90\% of
the (quasi-circular) {\it Fermi}-LAT error region \citep[7.7' radius, 95\%
c.l.;][]{Nolan12}.
No source is detected in the summed image, and we place a 90\%
c.l. upper limit on the 0.5--10~keV luminosity of
L$_\mathrm{X}$$<$2.2$\times 10^{31}$~erg~s$^{-1}$
(Table~\ref{table:redbacks}).
This upper limit uses the total XRT exposure at the center of the LAT
location (3.2~ksec). Over $\sim$40\% of the LAT error circle, the
effective XRT exposure gradually decreases, down to $\sim$0.5~ksec. If
we use this exposure instead (i.e., if the X-ray counterpart is near
the edge of the LAT location), the L$_\mathrm{X}$ upper limit becomes
$\sim 10^{32}$~erg~s$^{-1}$.

Figure~\ref{fig:pulsar} shows the upper limit on the luminosity of
J1628 in the pulsar state, together with the other 7 redbacks studied
in this work and M28-I.
If the X-ray counterpart is near the center of the LAT location, J1628
is the redback with the lowest L$_\mathrm{X}$.
Deeper observations covering a slightly wider field are needed to
identify the X-ray counterpart.
In any case our results show that, if the LAT location is accurate,
the luminosity of J1628 in the pulsar state is below
10$^{32}$~erg~s$^{-1}$

\subsection{PSR~J1723-2837}
\label{sec:j1723}

J1723 is somewhat peculiar among nearby redbacks, in the sense that it
was first discovered as a MRP (with P$_\mathrm{orb}$$\simeq$0.62~d and
P$_\mathrm{s}$$\simeq$1.86~ms) with the Parkes Multibeam survey
\citep[][who also reported optical photometric and spectroscopic
  studies]{Crawford13}, and then searched and found in the X-ray and
$\gamma$-ray bands \citep{Hui14,Bogdanov14b}.

The source is clearly detected in the two {\it Swift}-XRT
observations, at an average L$_\mathrm{X}$=[2.4$\pm$0.6]$\times
10^{32}$~erg~s$^{-1}$ and $\Gamma$=0.9$\pm$0.2.
Our {\it Swift} results are consistent with the average L$_\mathrm{X}$
and $\Gamma$ values measured with {\it Chandra} and {\it XMM} by
\citet[][who also find orbital modulation in
  L$_\mathrm{X}$]{Bogdanov14b}.

As can be seen in Figure~\ref{fig:pulsar} (right panel), the X-ray
properties of J1723 in the pulsar state (namely, L$_\mathrm{X}$ and
$\Gamma$) are very similar to those of the three redbacks that have
shown accretion: J1023, J12270 and M28-I.
For this reason, we suggest that J1723 is a good candidate to develop
active accretion and transition to the disk state in the near future
(see Sec.~\ref{sec:pulsar} for further discussion).

\subsection{PSR~J1816+4510}
\label{sec:j1816}

J1816 was discovered by \citet{Kaplan12} as a
P$_\mathrm{orb}$$\simeq$0.36~d, P$_\mathrm{s}$$\simeq$3.19~ms MRP, in
a Green Bank Telescope (GBT) observation targeted at the {\it Fermi}
source 2FGL~J1816.5+4511 (1FGL~J1816.7+4509).
The discovery and study of the optical and UV counterparts led to a
sub-arcsecond location and a lower limit on the neutron star mass of
1.8~M$_\odot$ \citep{Kaplan12,Kaplan13}.

\begin{figure*}
  \begin{center}
  \resizebox{1.8\columnwidth}{!}{\rotatebox{-90}{\includegraphics[]{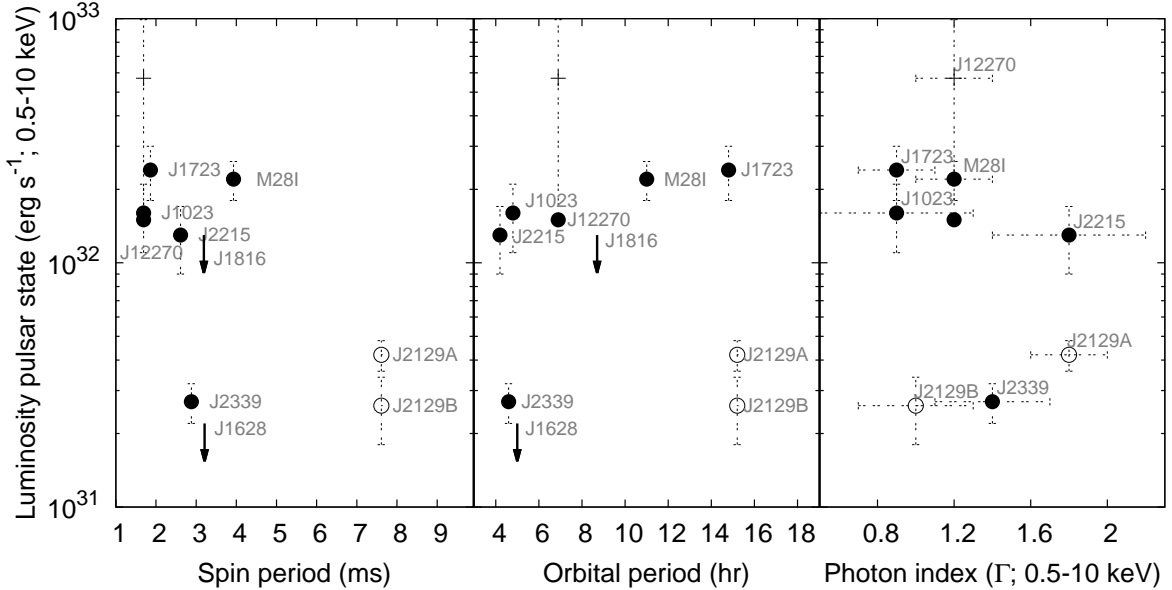}}}
  \caption{
Luminosity of the pulsar state in the X-ray band (0.5--10~keV),
plotted against P$_\mathrm{s}$ ({\it left}), P$_\mathrm{orb}$ ({\it
  middle}) and $\Gamma$ ({\it right}).
Sources are labeled and include all (eight) nearby redback MRPs
studied with {\it Swift} in this work, as well as M28-I for comparison
(Table~\ref{table:redbacks}).
We plot the two candidate counterparts to J2129 with open circles
(Sec.~\ref{sec:j2129}).
} %
    \label{fig:pulsar}
 \end{center}
\end{figure*}

We do not detect the source in the total 4.7~ksec {\it Swift}-XRT
exposure, summing all four available observations, and we place an
upper limit of L$_\mathrm{X}$$<$1.3$\times 10^{32}$~erg~s$^{-1}$
\citep[using the revised D=4.5~kpc reported
  by][]{Kaplan13}\footnote{In a similar study of the first observation
  (2.8~ksec), \citet{Kaplan12} report a 2$\sigma$ c.l. flux upper
  limit $\sim$18 times lower, which seems underestimated (their limit
  corresponds to a count rate of about 5$\times$10$^{-5}$~c~s$^{-1}$
  but the XRT background rate is typically $\sim$10$\times$ higher,
  see Sec.~\ref{sec:data}).}.
Figure~\ref{fig:pulsar} shows our {\it Swift}-XRT limit on the
luminosity of J1816 in the pulsar state.
J1816 has a pulsar state L$_\mathrm{X}$ below
$\sim$10$^{32}$~erg~s$^{-1}$, i.e., lower than the redbacks which have
shown accretion episodes.

\subsection{PSR~J2129--0428}
\label{sec:j2129}

J2129 was also discovered as a MRP with GBT \citep[][with
  P$_\mathrm{orb}$$\simeq$0.64~d and
  P$_\mathrm{s}$$\simeq$7.62~ms]{Hessels11,Ray12}, in targeted
observations of a LAT source (2FGL~J2129.8-0428/1FGL~J2129.8-0427).
No X-ray counterpart has been published to date\footnote{Roberts et
  al. presented a preliminary X-ray light curve, online at
  \url{http://aspen13.phys.wvu.edu/aspen\_talks/Roberts\_spiders.pdf},
  but we could not find the corresponding publication.}.
Optical studies are in progress \citep{Bellm13}, but no accurate
location has been published at the moment of writing.

The available {\it Swift}-XRT observation (totalling 10.4~ksec) covers
the entire (quasi-circular) {\it Fermi}-LAT error region \citep[8.6'
  radius, 95\% c.l.,][]{Nolan12}.
We detect two point sources inside the LAT error circle, which we name
J2129A and J2129B (in order of decreasing brightness).
We find the following UVOT-enhanced XRT position for J2129A: RA=21h
30m 08.16s, DEC=$-$04$^\circ$ 34' 53.8'' (J2000; with a 90\% error
radius of 2.3''), which is 7.2' from the LAT position.
For J2129B, the UVOT-enhanced XRT position is: RA=21h 29m 45.29s,
DEC=$-$04 $^\circ$ 29' 11.9'' (J2000; with a 90\% error radius of
5.8''), i.e., 1.7' from the LAT position.
Given the relatively large LAT error radius (8.6'), we consider both
J2129A and J2129B candidates for the X-ray counterpart of J2129 (and
we plot them using open circles in Figs.~\ref{fig:lg} and
\ref{fig:pulsar}).

Both X-ray counterpart candidates have relatively low luminosity,
L$_\mathrm{X} \lesssim$4$\times$10$^{31}$~erg~s$^{-1}$ (see
Table~\ref{table:redbacks} and Figure~\ref{fig:pulsar}). J2129A is
softer ($\Gamma$=1.8$\pm$0.2) than J2129B ($\Gamma$=1.0$\pm$0.3).
The X-ray luminosity and spectral shape of both candidates are
consistent with the rest of redbacks in the pulsar state.
Follow-up multi-wavelength observations are needed to identify which
of our two candidates is the true counterpart to J2129.
In any case, our results show that the X-ray luminosity of J2129 in
the pulsar state is low, well below 10$^{32}$~erg~s$^{-1}$.

\subsection{PSR~J2215+5135}
\label{sec:j2215}

Discovered in radio searches of the LAT source 1FGL~J2216.1+5139
(2FGL~J2215.7+5135), J2215 is a 2.61~ms MRP with the shortest orbital
period among Galactic field redbacks, P$_\mathrm{orb}$$\simeq$0.17~d
\citep{Hessels11}.
Optical \citep{Breton13} and X-ray \citep{Gentile14} studies have
identified, respectively, an irradiated companion star producing a
characteristic orbital light curve and a hard, variable X-ray
counterpart.

J2215 is detected in the two {\it Swift}-XRT observations, taken
about 3.5 years apart, with no evidence of variability between the two
epochs.
We find L$_\mathrm{X}$=[1.3$\pm$0.4]$\times 10^{32}$~erg~s$^{-1}$ and
$\Gamma$=1.8$\pm$0.4 \citep[Table~\ref{table:redbacks}; both
 consistent with the values reported by][]{Gentile14}.
Thus J2215 features a relatively high luminosity in the pulsar state,
similar to what we find for the three MRP-LMXB transition pulsars and
for J1723 (Figure~\ref{fig:pulsar}).
This suggests that J2215 is also a good candidate to show accretion
episodes and transitions to the disk and outburst states in the
future (Sec.~\ref{sec:pulsar}).

\subsection{PSR~J2339-0533}
\label{sec:j2339}

The discovery of J2339 followed a different path: the LAT source
0FGL~J2339.8--0530 was first identified as a likely MRP in a compact
orbit thanks to optical photometric and spectroscopic observations,
which revealed a P$_\mathrm{orb}$$\simeq$0.19~d
\citep{Kong11,Romani11,Kong12}.
More recently, radio and $\gamma$-ray pulsations have been discovered,
confirming the system as a redback with P$_\mathrm{s}$$\simeq$2.88~ms
\citep{Ray14}.

J2339 was observed 15 times with {\it Swift}-XRT, in 2009, 2011 and
2012, at similar XRT count rate levels.
We measure an average luminosity L$_\mathrm{X}$=[2.7$\pm$0.5]$\times
10^{31}$~erg~s$^{-1}$ and $\Gamma$=1.4$\pm$0.3 \citep[consistent with
  the values reported from a 2009 {\it Chandra} observation
  by][]{Romani11}.
We therefore find that J2339 is one of the four redbacks that feature
low L$_\mathrm{X}$ in the pulsar state (well below
10$^{32}$~erg~s$^{-1}$).

\section{The Disk State}
\label{sec:disk}

All three redbacks which have shown accretion episodes (J1023, J12270
and M28-I, also referred to as MRP-LMXB transition systems) have now
shown mode switching in the disk state: rapid ($\lesssim$500~s)
transitions between disk-active and disk-passive states with a factor
5--7 change in L$_\mathrm{X}$ and a purely non-thermal X-ray spectrum
\citep[see Table~\ref{table:redbacks},
  Figs.~\ref{fig:hist}--\ref{fig:lg} and][]{Linares14}.
This strongly suggests that X-ray mode switching is a common
phenomenon to all redbacks when they reach the accretion disk
state, at L$_\mathrm{X}$$\sim$10$^{33}$~erg~s$^{-1}$.
Having multiple redback MRPs observed in the disk state, subtle
differences begin to emerge. For instance, we find that J12270 is
significantly harder than J1023 in both the disk-active and
disk-passive states ($\Gamma$=1.15--1.36 vs. 1.62). This can be seen
in Figures~\ref{fig:spec} and \ref{fig:lg}.

It is worth noting that this behavior is so far unique to MRP-LMXB
transition systems.
To our knowledge, similar X-ray mode switching has never been observed
in neutron star transients in general \citep[transient LMXBs with no
  radio pulsar detected, such as Cen~X-4 or 4U~1608-52;
  e.g.,][]{Bernardini13} nor accreting millisecond X-ray pulsars in
particular (the subset of neutron star transients showing coherent
accretion-powered X-ray pulsations, such as SAX~J1808.4-3658 or
Aql~X-1), even though they have been observed at similar
L$_\mathrm{X}$
\citep[e.g.,][]{Campana97,Campana98b,Degenaar12c,Coti14}.
This implies a fundamental difference in the accretion flow-neutron
star interaction.

As noted in \citet{Linares14}, the two states found by
  \citet{Campana08} in SAX~J1808.4-3658 have different properties than
  the disk-active and disk-passive states of redbacks: i) they
  alternate on longer timescales (1--12~d), ii) they have lower
  luminosities (L$_\mathrm{X}$$<$5$\times$10$^{32}$~erg~s$^{-1}$) and
  iii) they show a clear spectral change ($\Gamma$=1.7--2.7).
More sensitive X-ray observations have not detected rapid transitions
between two clearly defined modes
\citep[e.g.,][]{Campana02,Wijnands03c}.
Furthermore, we apply the same analysis described in
Sec.~\ref{sec:data} to all available (36) {\it Swift}-XRT PC mode
observations, extended to time bins in the 200--2000~s range, and we do
not find evidence for a bimodal count rate distribution in the light
curves of SAX~J1808.4-3658.

The disk state of redback MRPs constitutes an intermediate stage
between the (radio) pulsar state
(L$_\mathrm{X}$$\sim$10$^{31}$--10$^{32}$~erg~s$^{-1}$) and a full
accretion outburst \citep[L$_\mathrm{X}$$>$10$^{34}$~erg~s$^{-1}$,
  seen so far only in M28-I,][]{Papitto13b}.
The non-detection of X-ray pulsations or a strong thermal
  component during the disk state, together with its relatively low
  L$_\mathrm{X}$, suggests that the accretion flow in this state does
  not reach the neutron star surface.
However, optical spectroscopy during the disk state reveals conclusive
evidence of an optically thick accretion disk: broad double-peaked
emission lines \citep{Wang09,deMartino13b,Linares14b}.
The disk states of redbacks last long (at least 6~yr and 1~yr in
J12270 and J1023, respectively), and therefore a stable disk solution
must exist at L$_\mathrm{X}$$\sim$10$^{33}$~erg~s$^{-1}$.

It is interesting to study where and how this accretion disk is
truncated, and for that direct observational constraints on the inner
disk radius (R$_\mathrm{in}$) are highly desirable.
A rough estimate of R$_\mathrm{in}$ can be obtained from the
magnetospheric radius,
R$_\mathrm{M}$=$7.8\left[B/10^8{\rm\,G}\right]^{4/7}\left[L_{bol}/L_{Edd}\right]^{-2/7}$~km
\citep[the radius where magnetic and ram pressures
    balance; e.g.,][we use here an Eddington luminosity of
  L$_\mathrm{Edd}$$\equiv$2.5$\times$10$^{38}$~erg~s$^{-1}$]{Lamb73,Psaltis99c,Patruno09}.
This implicitly assumes that the disk is truncated by the magnetic
field of the neutron star, and a linear relation between bolometric
luminosity L$_\mathrm{bol}$ and mass accretion rate in the disk
($\dot{M}$).

\begin{figure}
  \begin{center}
  \resizebox{0.95\columnwidth}{!}{\rotatebox{0}{\includegraphics[]{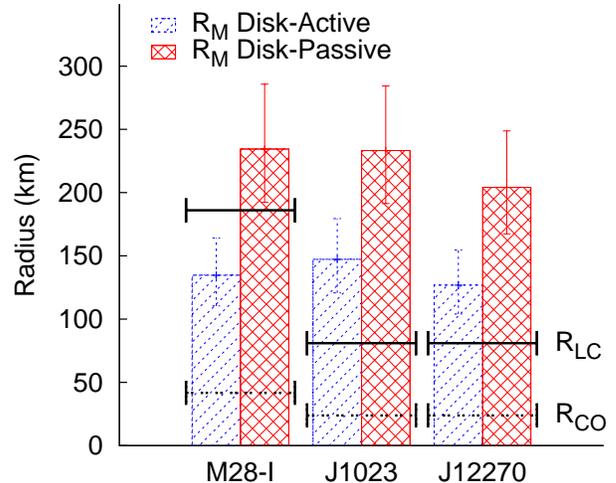}}}
  \caption{
Inferred magnetospheric radius (R$_\mathrm{M}$) for the disk-active
(blue dashed histograms) and disk-passive (red crossed histograms)
states of the three redbacks which have shown accretion (as labeled;
see Sec.~\ref{sec:disk} for R$_\mathrm{M}$ definition).
Horizontal solid and dashed black lines show the light cylinder radius
(R$_\mathrm{LC}$) and the corotation radius (R$_\mathrm{CO}$),
respectively.
Error bars show the effect on R$_\mathrm{M}$ of increasing/decreasing
the mass accretion rate by a factor 2.
} %
    \label{fig:disk}
 \end{center}
\end{figure}

Figure~\ref{fig:disk} shows the values of R$_\mathrm{M}$ inferred for
M28-I, J1023 and J12270 in their disk-active/passive states, compared
to the corresponding corotation (R$_\mathrm{CO}$) and light cylinder
(R$_\mathrm{LC}$) radii.
We correct L$_\mathrm{X}$ with a bolometric factor 3 following
\citet{Linares14} and use a value of B=10$^8$~G for the neutron star
magnetic field \citep[in accordance with the available measurements;
  see, e.g.,][for J1023]{Archibald13}.

In the ``tug-of-war'' scenario proposed by \citet{Linares14} for
M28-I, R$_\mathrm{M}$ during the disk-active state lies just inside
the light cylinder (i.e., R$_\mathrm{M}$$\lesssim$R$_\mathrm{LC}$),
and accretion proceeds down to the neutron star magnetosphere
\citep{Campana98}.
Fluctuations in $\dot{M}$ move R$_\mathrm{M}$ outside R$_\mathrm{LC}$,
which allows the radio pulsar to turn on.
This triggers a transition to the disk-passive state, where the lower
L$_\mathrm{X}$ is produced by a shock between the pulsar wind and the
innermost accretion flow.
A slight increase in $\dot{M}$ would then push the inner edge of the
disk back inside the light cylinder and turn off the radio pulsar
mechanism.
Thus in this simple qualitative model, supported by the inferred
R$_\mathrm{M}$ in M28-I (Fig.~\ref{fig:disk}), the X-ray mode switches
reflect a dynamic balance between the accretion flow and the pulsar
wind.

Let us now review this tug-of-war scenario in light of our new results
on the disk state of J1023 and J12270.
The light cylinder in J1023 and J12270 is a factor 2.3 smaller, due to
their 2.3 times shorter P$_\mathrm{s}$.
Given the weak scaling of R$_\mathrm{M}$ with $\dot{M}$ (R$_\mathrm{M}
\propto \dot{M}^{-2/7}$), L$_\mathrm{X}$ should increase by a factor
of almost 20
to compensate for the factor 2.3 faster spin and bring R$_\mathrm{M}$
back inside the light cylinder.
However, the L$_\mathrm{X}$ of J1023, J12270 (for D=1.4~kpc) and M28-I
in the active/passive states are consistent within a factor $<$1.7
(Table~\ref{table:redbacks}).
Therefore, we find that there is no strong dependence of the disk
state luminosity (and the inferred R$_\mathrm{in}$) on the pulsar's
spin or orbital period (P$_\mathrm{orb}$ in J1023 is also about 2.3
times shorter than M28-I).
This argues against the tug-of-war model in its simplest form: as
shown in Figure~\ref{fig:disk}, R$_\mathrm{M}$ does not cross the
light cylinder in the case of J1023 and J12270 when going from the
disk-active to the disk-passive state.

The R$_\mathrm{M}$ shown in Figure~\ref{fig:disk} suggest that
  J1023 and J12270 (with R$_\mathrm{M}$$>$R$_\mathrm{LC}$) may also
  harbor a rotation-powered pulsar in the disk state (both active and
  passive).
However, we stress that a more detailed physical model of the
transition between accretion and rotation power is needed to confirm
this, one that goes beyond the magnetospheric radius prescription.
Extensive radio observations of the disk-passive state can help
understand its nature, assessing whether or not the rotation-powered
pulsar is on.

Forces other than magnetic, such as radiation pressure from the radio
pulsar, are expected to act upon the inner accretion disk.
\citet{Eksi05} studied the truncation and stability of accretion disks
around radio pulsars.
They found that the radiation pressure of a rotation-powered pulsar
does not necessarily disrupt a surrounding disk, and therefore an
accretion disk and an active MRP can coexist.
According to their model, stable disk solutions exist with
R$_\mathrm{in}$$>$R$_\mathrm{LC}$, up to more than ten times
R$_\mathrm{LC}$ if the neutron star's magnetic inclination (angle
between spin and magnetic axes) is small \citep{Eksi05}.
In this context, we note that our highest estimate of the
magnetospheric radius in the disk-active state,
R$_\mathrm{M}$$\simeq$147~km~$\simeq$1.8$\times$R$_\mathrm{LC}$ in
J1023, only has stable solutions for magnetic inclinations lower than
15$^\circ$ \citep{Eksi05}.
This might indicate that redbacks have nearly aligned rotating
dipoles, although more detailed models tailored to the disk state of
redback MRPs are required to make quantitative statements.
In any case, our results strongly support the presence of a stable
accretion disk around redback millisecond pulsars with
R$_\mathrm{in}$$>$R$_\mathrm{LC}$.
Hence this new class of pulsars holds great potential to study
accretion flows near the light cylinder.

We measure an X-ray flux of 0.6~mCrab in the brightest disk-active
state in our sample (J12270), i.e., more than ten times fainter than
the 1-day sensitivity limits of currently active all-sky X-ray
monitors ({\it MAXI}: 8~mCrab at 5$\sigma$, \citealt{Matsuoka09}; {\it
  Swift}-BAT: 16~mCrab at 3$\sigma$, \citealt{Krimm13}).
Longer exposures \citep{Hiroi13,Krimm13} or improved all-sky monitors,
such as the Wide-Field Monitor onboard the proposed {\it LOFT}
mission\footnote{\url{http://sci.esa.int/loft/53447-loft-yellow-book/}},
may be able to detect new transitions of redbacks to the disk state.
Until then, {\it Swift}-XRT monitoring provides the most efficient
way of catching state transitions in nearby redback millisecond
pulsars.

An interesting question for future study is when exactly (at which
L$_\mathrm{X}$) accretion-powered X-ray pulsations appear, and whether
or not this coincides with the complete disappearance of
rotation-powered radio pulsations.
\citet{Papitto13b} set an upper limit of 17\% on the pulsed fraction
of M28-I at L$_\mathrm{X}$$\sim$10$^{33}$~erg~s$^{-1}$.
The redbacks compiled in this work offer the best prospects in this
respect, and given their short D and low N$_\mathrm{H}$ they should
allow a much more sensitive search for X-ray pulsations at low
L$_\mathrm{X}$. For instance, the collected X-ray count rates in the
disk-active state of J12270 (4.8 and 6.4~c~s$^{-1}$ with {\it XMM} and
the approved {\it NICER} mission,
respectively\footnote{\url{http://heasarc.gsfc.nasa.gov/Tools/w3pimms.html}})
are 25 times higher than those for M28-I at the same L$_\mathrm{X}$.

\section{The Pulsar State}
\label{sec:pulsar}

By studying the full sample of redbacks selected for this work, we
find that the X-ray luminosity of the pulsar state is in the
10$^{31}$--10$^{32}$~erg~s$^{-1}$ range, always lower than that of the
disk state (Fig.~\ref{fig:lg}).
Moreover, redbacks can be roughly divided in two groups based on their
luminosity in the pulsar state (Figure~\ref{fig:pulsar}): those having
relatively high L$_\mathrm{X}$ ($\gtrsim$10$^{32}$~erg~s$^{-1}$:
M28-I, J1023, J12270, J1723 and J2215) and those having low
L$_\mathrm{X}$ ($<$10$^{32}$~erg~s$^{-1}$: J1628, J2339, J2129 and
probably J1816).
The XRT spectra of the pulsar state are hard, with $\Gamma$ between
0.9 and 1.8.
We do not find strong dependence of L$_\mathrm{X}$ or $\Gamma$ on
P$_\mathrm{s}$, P$_\mathrm{orb}$ or M$_\mathrm{C}$.

Based on their high pulsar state luminosities, we suggest that J1723
and J2215 are good candidates to show accretion episodes in the near
future (Secs.~\ref{sec:j1723} and \ref{sec:j2215}).
A higher L$_\mathrm{X}$ may indicate stronger interaction with the
companion star, due to a higher spin-down luminosity or to a larger
fraction of the pulsar wind being shocked. More intense ablation of
the companion's heated face could result into enhanced mass transfer,
and potentially feed accretion disk episodes such as those seen in
J1023 and J12270.
More sensitive X-ray missions are better suited to study in detail
this state \citep{Bogdanov11b,Bogdanov14}, although we have shown that
{\it Swift}-XRT can detect nearby redbacks in the pulsar state with
moderate exposure times ($\sim$10~ksec), down to just a few times
10$^{31}$~erg~s$^{-1}$.
Measurements of the radio pulsar spin-down rate may be possible soon
for more redbacks, and together with the luminosities reported herein
will help constrain the energetics of the pulsar wind shock.

\section{Summary and Conclusions}
\label{sec:conclusions}

Redbacks have provided the long-sought link between MRPs and LMXBs,
and we are in the process of understanding their properties as a
class.
They allow us to study in an unprecented way the transition between
accretion and rotation power around neutron stars, as well as the
interaction between millisecond pulsars and accretion flows.

We have presented the first systematic X-ray study of redback
millisecond pulsars, using more than 100 {\it Swift} observations of
eight nearby systems and covering luminosities between 10$^{31}$ and
10$^{34}$~erg~s$^{-1}$.
We next summarize our main results and conclusions.

\begin{itemize}

\item 

The X-ray luminosity of redbacks can vary by six orders of magnitude,
defining three main states: i) {\it outburst state}
(10$^{34}$$<$L$_\mathrm{X}$$<$10$^{37}$~erg~s$^{-1}$), seen only in one
system to date; ii) {\it disk state}
(4$\times$10$^{32}$$<$L$_\mathrm{X}$$<$10$^{34}$~erg~s$^{-1}$), observed
in three systems to date; and iii) {\it pulsar state}
(10$^{31}<$L$_\mathrm{X}<4\times$10$^{32}$~erg~s$^{-1}$), where at
least seven redbacks have been detected in X-rays.

\item

All redbacks which have transitioned to the disk state (J1023, J12270
and M28-I) have shown {\it X-ray mode switching}: fast back-and-forth
transitions between disk-active and disk-passive states, with a factor
5--7 change in luminosity.
This strongly suggests that mode switching, which has not been
observed in ``normal'' LMXBs, is a universal phenomenon of redbacks in
the disk state.

\item 

The current sample of redbacks can be roughly divided in two groups
according to their luminosity in the pulsar state (bright or faint,
above or below 10$^{32}$~erg~s$^{-1}$).
All three redbacks which have shown evidence for accretion (J1023,
J12270 and M28-I), have L$_\mathrm{X}$$>$10$^{32}$~erg~s$^{-1}$.
Based on this we suggest that J1723 and J2215 are promising candidates
to develop accretion disks.

\end{itemize}

\footnotetext{After this work was submitted, \citet{Tendulkar14}
  published results from a {\it NuStar} hard X-ray study of
  J1023. They find two distinct states in the October 2013 observation
  of J1023 (which they call dip and non-dip states) with timescales,
  luminosity and spectral shape similar to the disk-passive and
  disk-active states that we report and discuss herein.}

\textbf{Acknowledgments:}

I am most grateful to A. Alpar and J. Homan for their insightful
comments on the manuscript, to M. Roberts for a conversation that
sparked this work and to S. Bogdanov for sharing his XMM results on
J1023.
I also thank the anonymous referee for constructive comments that
improved the paper.
This research has made use of data obtained from NASA's High Energy
Astrophysics Science Archive Research Center (HEASARC), and data
supplied by the UK Swift Science Data Center at the University of
Leicester.


\begin{thebibliography}{76}
\expandafter\ifx\csname natexlab\endcsname\relax\def\natexlab#1{#1}\fi

\bibitem[{{Abdo} {et~al}\mbox{.}(2010){Abdo}, {Ackermann}, {Ajello},
  {Allafort}, {Antolini}, {Atwood}, {Axelsson}, {Baldini}, {Ballet},
  {Barbiellini}, \& et~al.}]{Abdo10}
{Abdo} A.~A. {et~al.}, 2010, \apjs, 188, 405

\bibitem[{{Alpar} {et~al}\mbox{.}(1982){Alpar}, {Cheng}, {Ruderman}, \&
  {Shaham}}]{Alpar82}
{Alpar} M.~A., {Cheng} A.~F., {Ruderman} M.~A., {Shaham} J., 1982, \nat, 300,
  728

\bibitem[{{Archibald} {et~al}\mbox{.}(2010){Archibald}, {Kaspi}, {Bogdanov},
  {Hessels}, {Stairs}, {Ransom}, \& {McLaughlin}}]{Archibald10}
{Archibald} A.~M., {Kaspi} V.~M., {Bogdanov} S., {Hessels} J.~W.~T., {Stairs}
  I.~H., {Ransom} S.~M., {McLaughlin} M.~A., 2010, \apj, 722, 88

\bibitem[{{Archibald} {et~al}\mbox{.}(2013){Archibald}, {Kaspi}, {Hessels},
  {Stappers}, {Janssen}, \& {Lyne}}]{Archibald13}
{Archibald} A.~M., {Kaspi} V.~M., {Hessels} J.~W.~T., {Stappers} B., {Janssen}
  G., {Lyne} A., 2013, Submitted to ApJ; ArXiv 1311.5161

\bibitem[{{Archibald} {et~al}\mbox{.}(2009){Archibald}, {Stairs}, {Ransom},
  {Kaspi}, {Kondratiev}, {Lorimer}, {McLaughlin}, {Boyles}, {Hessels}, {Lynch},
  {van Leeuwen}, {Roberts}, {Jenet}, {Champion}, {Rosen}, {Barlow}, {Dunlap},
  \& {Remillard}}]{Archibald09}
{Archibald} A.~M. {et~al.}, 2009, Science, 324, 1411

\bibitem[{{Arnaud}(1996)}]{Arnaud96}
{Arnaud} K.~A., 1996, in Astronomical Society of the Pacific Conference Series,
  Vol. 101, Astronomical Data Analysis Software and Systems V, {Jacoby} G.~H.,
  {Barnes} J., eds., pp. 17--+

\bibitem[{{Arons} \& {Tavani}(1993)}]{Arons93}
{Arons} J., {Tavani} M., 1993, \apj, 403, 249

\bibitem[{{Bassa} {et~al}\mbox{.}(2014){Bassa}, {Patruno}, {Hessels}, {Keane},
  {Monard}, {Mahony}, {Bogdanov}, {Corbel}, {Edwards}, {Archibald}, {Janssen},
  {Stappers}, \& {Tendulkar}}]{Bassa14}
{Bassa} C.~G. {et~al.}, 2014, \mnras, 441, 1825

\bibitem[{{Begin}(2006)}]{Begin06}
{Begin} S., 2006, Master's thesis, Faculty of Physics, UBC

\bibitem[{{Bellm} {et~al}\mbox{.}(2013){Bellm}, {Djorgovski}, {Drake},
  {Hessels}, {Kulkarni}, {Levitan}, {Mahabal}, {Phinney}, {Ransom}, {Roberts},
  {Prince}, {Sesar}, \& {Tang}}]{Bellm13}
{Bellm} E. {et~al.}, 2013, in American Astronomical Society Meeting Abstracts,
  Vol. 221, American Astronomical Society Meeting Abstracts, p. 154.10

\bibitem[{{Bernardini} {et~al}\mbox{.}(2013){Bernardini}, {Cackett}, {Brown},
  {D'Angelo}, {Degenaar}, {Miller}, {Reynolds}, \& {Wijnands}}]{Bernardini13}
{Bernardini} F., {Cackett} E.~M., {Brown} E.~F., {D'Angelo} C., {Degenaar} N.,
  {Miller} J.~M., {Reynolds} M., {Wijnands} R., 2013, \mnras, 436, 2465

\bibitem[{{Bogdanov} {et~al}\mbox{.}(2011{\natexlab{a}}){Bogdanov},
  {Archibald}, {Hessels}, {Kaspi}, {Lorimer}, {McLaughlin}, {Ransom}, \&
  {Stairs}}]{Bogdanov11b}
{Bogdanov} S., {Archibald} A.~M., {Hessels} J.~W.~T., {Kaspi} V.~M., {Lorimer}
  D., {McLaughlin} M.~A., {Ransom} S.~M., {Stairs} I.~H., 2011{\natexlab{a}},
  \apj, 742, 97

\bibitem[{{Bogdanov} {et~al}\mbox{.}(2014{\natexlab{a}}){Bogdanov}, {Esposito},
  {Crawford}, {Possenti}, {McLaughlin}, \& {Freire}}]{Bogdanov14b}
{Bogdanov} S., {Esposito} P., {Crawford}, III F., {Possenti} A., {McLaughlin}
  M.~A., {Freire} P., 2014{\natexlab{a}}, \apj, 781, 6

\bibitem[{{Bogdanov}, {Grindlay} \& {van den Berg}(2005){Bogdanov}, {Grindlay},
  \& {van den Berg}}]{Bogdanov05}
{Bogdanov} S., {Grindlay} J.~E., {van den Berg} M., 2005, \apj, 630, 1029

\bibitem[{{Bogdanov} {et~al}\mbox{.}(2014{\natexlab{b}}){Bogdanov}, {Patruno},
  {Archibald}, {Bassa}, {Hessels}, {Janssen}, \& {Stappers}}]{Bogdanov14}
{Bogdanov} S., {Patruno} A., {Archibald} A.~M., {Bassa} C., {Hessels} J.~W.~T.,
  {Janssen} G.~H., {Stappers} B.~W., 2014{\natexlab{b}}, Submitted to ApJ;
  ArXiv 1402.6324

\bibitem[{{Bogdanov} {et~al}\mbox{.}(2011{\natexlab{b}}){Bogdanov}, {van den
  Berg}, {Servillat}, {Heinke}, {Grindlay}, {Stairs}, {Ransom}, {Freire},
  {B{\'e}gin}, \& {Becker}}]{Bogdanov11}
{Bogdanov} S. {et~al.}, 2011{\natexlab{b}}, \apj, 730, 81

\bibitem[{{Breton} {et~al}\mbox{.}(2013){Breton}, {van Kerkwijk}, {Roberts},
  {Hessels}, {Camilo}, {McLaughlin}, {Ransom}, {Ray}, \& {Stairs}}]{Breton13}
{Breton} R.~P. {et~al.}, 2013, \apj, 769, 108

\bibitem[{{Brown}, {Bildsten} \& {Rutledge}(1998){Brown}, {Bildsten}, \&
  {Rutledge}}]{Brown98}
{Brown} E.~F., {Bildsten} L., {Rutledge} R.~E., 1998, \apjl, 504, L95

\bibitem[{{Camilo} {et~al}\mbox{.}(2000){Camilo}, {Lorimer}, {Freire}, {Lyne},
  \& {Manchester}}]{Camilo00}
{Camilo} F., {Lorimer} D.~R., {Freire} P., {Lyne} A.~G., {Manchester} R.~N.,
  2000, \apj, 535, 975

\bibitem[{{Campana} {et~al}\mbox{.}(1998{\natexlab{a}}){Campana}, {Colpi},
  {Mereghetti}, {Stella}, \& {Tavani}}]{Campana98}
{Campana} S., {Colpi} M., {Mereghetti} S., {Stella} L., {Tavani} M.,
  1998{\natexlab{a}}, \aapr, 8, 279

\bibitem[{{Campana} {et~al}\mbox{.}(1997){Campana}, {Mereghetti}, {Stella}, \&
  {Colpi}}]{Campana97}
{Campana} S., {Mereghetti} S., {Stella} L., {Colpi} M., 1997, \aap, 324, 941

\bibitem[{{Campana} {et~al}\mbox{.}(2008){Campana}, {Panagia}, {Lazzati},
  {Beardmore}, {Cusumano}, {Godet}, {Chincarini}, {Covino}, {Della Valle},
  {Guidorzi}, {Malesani}, {Moretti}, {Perna}, {Romano}, \&
  {Tagliaferri}}]{Campana08}
{Campana} S. {et~al.}, 2008, \apjl, 683, L9

\bibitem[{{Campana} {et~al}\mbox{.}(2002){Campana}, {Stella}, {Gastaldello},
  {Mereghetti}, {Colpi}, {Israel}, {Burderi}, {Di Salvo}, \&
  {Robba}}]{Campana02}
{Campana} S. {et~al.}, 2002, \apjl, 575, L15

\bibitem[{{Campana} {et~al}\mbox{.}(1998{\natexlab{b}}){Campana}, {Stella},
  {Mereghetti}, {Colpi}, {Tavani}, {Ricci}, {Fiume}, \& {Belloni}}]{Campana98b}
{Campana} S., {Stella} L., {Mereghetti} S., {Colpi} M., {Tavani} M., {Ricci}
  D., {Fiume} D.~D., {Belloni} T., 1998{\natexlab{b}}, \apjl, 499, L65+

\bibitem[{{Coti Zelati} {et~al}\mbox{.}(2014){Coti Zelati}, {Campana},
  {D'Avanzo}, \& {Melandri}}]{Coti14}
{Coti Zelati} F., {Campana} S., {D'Avanzo} P., {Melandri} A., 2014, \mnras,
  438, 2634

\bibitem[{{Crawford} {et~al}\mbox{.}(2013){Crawford}, {Lyne}, {Stairs},
  {Kaplan}, {McLaughlin}, {Freire}, {Burgay}, {Camilo}, {D'Amico}, {Faulkner},
  {Kramer}, {Lorimer}, {Manchester}, {Possenti}, \& {Steeghs}}]{Crawford13}
{Crawford} F. {et~al.}, 2013, \apj, 776, 20

\bibitem[{{D'Amico} {et~al}\mbox{.}(2001){D'Amico}, {Possenti}, {Manchester},
  {Sarkissian}, {Lyne}, \& {Camilo}}]{DAmico01}
{D'Amico} N., {Possenti} A., {Manchester} R.~N., {Sarkissian} J., {Lyne} A.~G.,
  {Camilo} F., 2001, \apjl, 561, L89

\bibitem[{{de Martino} {et~al}\mbox{.}(2013{\natexlab{a}}){de Martino},
  {Belloni}, {Falanga}, {Papitto}, {Motta}, {Pellizzoni}, {Evangelista},
  {Piano}, {Masetti}, {Bonnet-Bidaud}, {Mouchet}, {Mukai}, \&
  {Possenti}}]{deMartino13}
{de Martino} D. {et~al.}, 2013{\natexlab{a}}, \aap, 550, A89

\bibitem[{{de Martino} {et~al}\mbox{.}(2013{\natexlab{b}}){de Martino},
  {Casares}, {Mason}, {Kotze}, {Buckley}, {Bonnet-Bidaud}, {Belloni},
  {Mouchet}, \& {Falanga}}]{deMartino13b}
{de Martino} D. {et~al.}, 2013{\natexlab{b}}, The Astronomer's Telegram, 5651,
  1

\bibitem[{{de Martino} {et~al}\mbox{.}(2010){de Martino}, {Falanga},
  {Bonnet-Bidaud}, {Belloni}, {Mouchet}, {Masetti}, {Andruchow}, {Cellone},
  {Mukai}, \& {Matt}}]{deMartino10}
{de Martino} D. {et~al.}, 2010, \aap, 515, A25

\bibitem[{{Degenaar} \& {Wijnands}(2012)}]{Degenaar12c}
{Degenaar} N., {Wijnands} R., 2012, \mnras, 422, 581

\bibitem[{{Deller} {et~al}\mbox{.}(2012){Deller}, {Archibald}, {Brisken},
  {Chatterjee}, {Janssen}, {Kaspi}, {Lorimer}, {Lyne}, {McLaughlin}, {Ransom},
  {Stairs}, \& {Stappers}}]{Deller12}
{Deller} A.~T. {et~al.}, 2012, \apjl, 756, L25

\bibitem[{{Ek{\c s}{\.I}} \& {Alpar}(2005)}]{Eksi05}
{Ek{\c s}{\.I}} K.~Y., {Alpar} M.~A., 2005, \apj, 620, 390

\bibitem[{{Evans} {et~al}\mbox{.}(2009){Evans}, {Beardmore}, {Page}, {Osborne},
  {O'Brien}, {Willingale}, {Starling}, {Burrows}, {Godet}, {Vetere}, {Racusin},
  {Goad}, {Wiersema}, {Angelini}, {Capalbi}, {Chincarini}, {Gehrels}, {Kennea},
  {Margutti}, {Morris}, {Mountford}, {Pagani}, {Perri}, {Romano}, \&
  {Tanvir}}]{Evans09}
{Evans} P.~A. {et~al.}, 2009, \mnras, 397, 1177

\bibitem[{{Ferrigno} {et~al}\mbox{.}(2014){Ferrigno}, {Bozzo}, {Papitto},
  {Rea}, {Pavan}, {Campana}, {Wieringa}, {Filipovi{\'c}}, {Falanga}, \&
  {Stella}}]{Ferrigno14}
{Ferrigno} C. {et~al.}, 2014, \aap, 567, A77

\bibitem[{{Fruchter}, {Stinebring} \& {Taylor}(1988){Fruchter}, {Stinebring},
  \& {Taylor}}]{Fruchter88}
{Fruchter} A.~S., {Stinebring} D.~R., {Taylor} J.~H., 1988, \nat, 333, 237

\bibitem[{{Gentile} {et~al}\mbox{.}(2014){Gentile}, {Roberts}, {McLaughlin},
  {Camilo}, {Hessels}, {Kerr}, {Ransom}, {Ray}, \& {Stairs}}]{Gentile14}
{Gentile} P.~A. {et~al.}, 2014, \apj, 783, 69

\bibitem[{{Goad} {et~al}\mbox{.}(2007){Goad}, {Tyler}, {Beardmore}, {Evans},
  {Rosen}, {Osborne}, {Starling}, {Marshall}, {Yershov}, {Burrows}, {Gehrels},
  {Roming}, {Moretti}, {Capalbi}, {Hill}, {Kennea}, {Koch}, \& {vanden
  Berk}}]{Goad07}
{Goad} M.~R. {et~al.}, 2007, \aap, 476, 1401

\bibitem[{{Harris}(1996)}]{Harris96}
{Harris} W.~E., 1996, \aj, 112, 1487

\bibitem[{{Heinke} {et~al}\mbox{.}(2006){Heinke}, {Rybicki}, {Narayan}, \&
  {Grindlay}}]{Heinke06b}
{Heinke} C.~O., {Rybicki} G.~B., {Narayan} R., {Grindlay} J.~E., 2006, \apj,
  644, 1090

\bibitem[{{Hessels} {et~al}\mbox{.}(2011){Hessels}, {Roberts}, {McLaughlin},
  {Ray}, {Bangale}, {Ransom}, {Kerr}, {Camilo}, \& {Decesar}}]{Hessels11}
{Hessels} J.~W.~T. {et~al.}, 2011, in American Institute of Physics Conference
  Series, Vol. 1357, American Institute of Physics Conference Series, {Burgay}
  M., {D'Amico} N., {Esposito} P., {Pellizzoni} A., {Possenti} A., eds., pp.
  40--43

\bibitem[{{Hill} {et~al}\mbox{.}(2011){Hill}, {Szostek}, {Corbel}, {Camilo},
  {Corbet}, {Dubois}, {Dubus}, {Edwards}, {Ferrara}, {Kerr}, {Koerding},
  {Kozie{\l}}, \& {Stawarz}}]{Hill11}
{Hill} A.~B. {et~al.}, 2011, \mnras, 415, 235

\bibitem[{{Hiroi} {et~al}\mbox{.}(2013){Hiroi}, {Ueda}, {Hayashida},
  {Shidatsu}, {Sato}, {Kawamuro}, {Sugizaki}, {Nakahira}, {Serino}, {Kawai},
  {Matsuoka}, {Mihara}, {Morii}, {Nakajima}, {Negoro}, {Sakamoto}, {Tomida},
  {Tsuboi}, {Tsunemi}, {Ueno}, {Yamaoka}, {Yoshida}, {Asada}, {Eguchi},
  {Hanayama}, {Higa}, {Ishikawa}, {Ishikawa}, {Isobe}, {Kohama}, {Kimura},
  {Morihana}, {Nakagawa}, {Nakano}, {Nishimura}, {Ogawa}, {Sasaki}, {Sugimoto},
  {Takagi}, {Usui}, {Yamamoto}, {Yamauchi}, \& {Yoshidome}}]{Hiroi13}
{Hiroi} K. {et~al.}, 2013, \apjs, 207, 36

\bibitem[{{Hui} {et~al}\mbox{.}(2014){Hui}, {Tam}, {Takata}, {Kong}, {Cheng},
  {Wu}, {Lin}, \& {Wu}}]{Hui14}
{Hui} C.~Y., {Tam} P.~H.~T., {Takata} J., {Kong} A.~K.~H., {Cheng} K.~S., {Wu}
  J.~H.~K., {Lin} L.~C.~C., {Wu} E.~M.~H., 2014, \apjl, 781, L21

\bibitem[{{Kalberla} {et~al}\mbox{.}(2005){Kalberla}, {Burton}, {Hartmann},
  {Arnal}, {Bajaja}, {Morras}, \& {P{\"o}ppel}}]{Kalberla05}
{Kalberla} P.~M.~W., {Burton} W.~B., {Hartmann} D., {Arnal} E.~M., {Bajaja} E.,
  {Morras} R., {P{\"o}ppel} W.~G.~L., 2005, \aap, 440, 775

\bibitem[{{Kaplan} {et~al}\mbox{.}(2013){Kaplan}, {Bhalerao}, {van Kerkwijk},
  {Koester}, {Kulkarni}, \& {Stovall}}]{Kaplan13}
{Kaplan} D.~L., {Bhalerao} V.~B., {van Kerkwijk} M.~H., {Koester} D.,
  {Kulkarni} S.~R., {Stovall} K., 2013, \apj, 765, 158

\bibitem[{{Kaplan} {et~al}\mbox{.}(2012){Kaplan}, {Stovall}, {Ransom},
  {Roberts}, {Kotulla}, {Archibald}, {Biwer}, {Boyles}, {Dartez}, {Day},
  {Ford}, {Garcia}, {Hessels}, {Jenet}, {Karako}, {Kaspi}, {Kondratiev},
  {Lorimer}, {Lynch}, {McLaughlin}, {Rohr}, {Siemens}, {Stairs}, \& {van
  Leeuwen}}]{Kaplan12}
{Kaplan} D.~L. {et~al.}, 2012, \apj, 753, 174

\bibitem[{{Kong} {et~al}\mbox{.}(2012){Kong}, {Huang}, {Cheng}, {Takata},
  {Yatsu}, {Cheung}, {Donato}, {Lin}, {Kataoka}, {Takahashi}, {Maeda}, {Hui},
  \& {Tam}}]{Kong12}
{Kong} A.~K.~H. {et~al.}, 2012, \apjl, 747, L3

\bibitem[{{Kong} {et~al}\mbox{.}(2011){Kong}, {Huang}, {Tam}, {Cheng},
  {Takata}, \& {Hui}}]{Kong11}
{Kong} A.~K.~H., {Huang} R.~H.~H., {Tam} P.~H.~T., {Cheng} K.~S., {Takata} J.,
  {Hui} C.~Y., 2011, in American Astronomical Society Meeting Abstracts \#218,
  p. 320.05

\bibitem[{{Krimm} {et~al}\mbox{.}(2013){Krimm}, {Holland}, {Corbet},
  {Pearlman}, {Romano}, {Kennea}, {Bloom}, {Barthelmy}, {Baumgartner},
  {Cummings}, {Gehrels}, {Lien}, {Markwardt}, {Palmer}, {Sakamoto},
  {Stamatikos}, \& {Ukwatta}}]{Krimm13}
{Krimm} H.~A. {et~al.}, 2013, \apjs, 209, 14

\bibitem[{{Lamb}, {Pethick} \& {Pines}(1973){Lamb}, {Pethick}, \&
  {Pines}}]{Lamb73}
{Lamb} F.~K., {Pethick} C.~J., {Pines} D., 1973, \apj, 184, 271

\bibitem[{{Linares} {et~al}\mbox{.}(2014{\natexlab{a}}){Linares}, {Bahramian},
  {Heinke}, {Wijnands}, {Patruno}, {Altamirano}, {Homan}, {Bogdanov}, \&
  {Pooley}}]{Linares14}
{Linares} M. {et~al.}, 2014{\natexlab{a}}, \mnras, 438, 251

\bibitem[{{Linares} {et~al}\mbox{.}(2014{\natexlab{b}}){Linares}, {Casares},
  {Rodriguez-Gil}, \& {Shahbaz}}]{Linares14b}
{Linares} M., {Casares} J., {Rodriguez-Gil} P., {Shahbaz} T.,
  2014{\natexlab{b}}, The Astronomer's Telegram, 5868, 1

\bibitem[{{Matsuoka} {et~al}\mbox{.}(2009){Matsuoka}, {Kawasaki}, {Ueno},
  {Tomida}, {Kohama}, {Suzuki}, {Adachi}, {Ishikawa}, {Mihara}, {Sugizaki},
  {Isobe}, {Nakagawa}, {Tsunemi}, {Miyata}, {Kawai}, {Kataoka}, {Morii},
  {Yoshida}, {Negoro}, {Nakajima}, {Ueda}, {Chujo}, {Yamaoka}, {Yamazaki},
  {Nakahira}, {You}, {Ishiwata}, {Miyoshi}, {Eguchi}, {Hiroi}, {Katayama}, \&
  {Ebisawa}}]{Matsuoka09}
{Matsuoka} M. {et~al.}, 2009, \pasj, 61, 999

\bibitem[{{Nolan} {et~al}\mbox{.}(2012){Nolan}, {Abdo}, {Ackermann}, {Ajello},
  {Allafort}, {Antolini}, {Atwood}, {Axelsson}, {Baldini}, {Ballet}, \&
  et~al.}]{Nolan12}
{Nolan} P.~L. {et~al.}, 2012, \apjs, 199, 31

\bibitem[{{Papitto} {et~al}\mbox{.}(2013){Papitto}, {Ferrigno}, {Bozzo}, {Rea},
  {Pavan}, {Burderi}, {Burgay}, {Campana}, {di Salvo}, {Falanga},
  {Filipovi{\'c}}, {Freire}, {Hessels}, {Possenti}, {Ransom}, {Riggio},
  {Romano}, {Sarkissian}, {Stairs}, {Stella}, {Torres}, {Wieringa}, \&
  {Wong}}]{Papitto13b}
{Papitto} A. {et~al.}, 2013, \nat, 501, 517

\bibitem[{{Papitto}, {Torres} \& {Li}(2014){Papitto}, {Torres}, \&
  {Li}}]{Papitto14}
{Papitto} A., {Torres} D.~F., {Li} J., 2014, \mnras, 438, 2105

\bibitem[{{Patruno} {et~al}\mbox{.}(2014){Patruno}, {Archibald}, {Hessels},
  {Bogdanov}, {Stappers}, {Bassa}, {Janssen}, {Kaspi}, {Tendulkar}, \&
  {Lyne}}]{Patruno14}
{Patruno} A. {et~al.}, 2014, \apjl, 781, L3

\bibitem[{{Patruno} {et~al}\mbox{.}(2009){Patruno}, {Watts}, {Klein Wolt},
  {Wijnands}, \& {van der Klis}}]{Patruno09}
{Patruno} A., {Watts} A., {Klein Wolt} M., {Wijnands} R., {van der Klis} M.,
  2009, \apj, 707, 1296

\bibitem[{{Possenti} {et~al}\mbox{.}(2003){Possenti}, {D'Amico}, {Manchester},
  {Camilo}, {Lyne}, {Sarkissian}, \& {Corongiu}}]{Possenti03}
{Possenti} A., {D'Amico} N., {Manchester} R.~N., {Camilo} F., {Lyne} A.~G.,
  {Sarkissian} J., {Corongiu} A., 2003, \apj, 599, 475

\bibitem[{{Psaltis} \& {Chakrabarty}(1999)}]{Psaltis99c}
{Psaltis} D., {Chakrabarty} D., 1999, \apj, 521, 332

\bibitem[{{Ray} {et~al}\mbox{.}(2012){Ray}, {Abdo}, {Parent}, {Bhattacharya},
  {Bhattacharyya}, {Camilo}, {Cognard}, {Theureau}, {Ferrara}, {Harding},
  {Thompson}, {Freire}, {Guillemot}, {Gupta}, {Roy}, {Hessels}, {Johnston},
  {Keith}, {Shannon}, {Kerr}, {Michelson}, {Romani}, {Kramer}, {McLaughlin},
  {Ransom}, {Roberts}, {Saz Parkinson}, {Ziegler}, {Smith}, {Stappers},
  {Weltevrede}, \& {Wood}}]{Ray12}
{Ray} P.~S. {et~al.}, 2012, 2011 Fermi Symposium proceedings - eConf C110509;
  ArXiv 1205.3089

\bibitem[{{Ray} {et~al}\mbox{.}(2014){Ray}, {Belfiore}, {Saz Parkinson},
  {Polisensky}, {Ransom}, {Romani}, {Hessels}, {Razzano}, {Bhattacharyya},
  {Roy}, {Cognard}, \& {Pulsar Search Consortium}}]{Ray14}
{Ray} P.~S. {et~al.}, 2014, in American Astronomical Society Meeting Abstracts,
  Vol. 223, American Astronomical Society Meeting Abstracts, p. 140.07

\bibitem[{{Roberts}(2011)}]{Roberts11}
{Roberts} M.~S.~E., 2011, in American Institute of Physics Conference Series
  (arXiv: 1103.0819), Vol. 1357, American Institute of Physics Conference
  Series, {Burgay} M., {D'Amico} N., {Esposito} P., {Pellizzoni} A., {Possenti}
  A., eds., pp. 127--130

\bibitem[{{Roberts}(2013)}]{Roberts13}
{Roberts} M.~S.~E., 2013, in IAU Symposium (arXiv:1210.6903), Vol. 291, IAU
  Symposium, pp. 127--132

\bibitem[{{Romani} \& {Shaw}(2011)}]{Romani11}
{Romani} R.~W., {Shaw} M.~S., 2011, \apjl, 743, L26

\bibitem[{{Roy}, {Bhattacharyya} \& {Ray}(2014){Roy}, {Bhattacharyya}, \&
  {Ray}}]{Roy14}
{Roy} J., {Bhattacharyya} B., {Ray} P., 2014, The Astronomer's Telegram, 5890,
  1

\bibitem[{{Stappers} {et~al}\mbox{.}(2014){Stappers}, {Archibald}, {Hessels},
  {Bassa}, {Bogdanov}, {Janssen}, {Kaspi}, {Lyne}, {Patruno}, {Tendulkar},
  {Hill}, \& {Glanzman}}]{Stappers14}
{Stappers} B.~W. {et~al.}, 2014, \apj, 790, 39

\bibitem[{{Stappers} {et~al}\mbox{.}(2003){Stappers}, {Gaensler}, {Kaspi}, {van
  der Klis}, \& {Lewin}}]{Stappers03}
{Stappers} B.~W., {Gaensler} B.~M., {Kaspi} V.~M., {van der Klis} M., {Lewin}
  W.~H.~G., 2003, Science, 299, 1372

\bibitem[{{Takata} {et~al}\mbox{.}(2014){Takata}, {Li}, {Leung}, {Kong}, {Tam},
  {Hui}, {Wu}, {Xing}, {Cao}, {Tang}, {Wang}, \& {Cheng}}]{Takata14}
{Takata} J. {et~al.}, 2014, \apj, 785, 131

\bibitem[{{Tam} {et~al}\mbox{.}(2010){Tam}, {Hui}, {Huang}, {Kong}, {Takata},
  {Lin}, {Yang}, {Cheng}, \& {Taam}}]{Tam10}
{Tam} P.~H.~T. {et~al.}, 2010, \apjl, 724, L207

\bibitem[{{Tauris}(2012)}]{Tauris12}
{Tauris} T.~M., 2012, Science, 335, 561

\bibitem[{{Tendulkar} {et~al}\mbox{.}(2014){Tendulkar}, {Yang}, {An}, {Kaspi},
  {Archibald}, {Bassa}, {Bellm}, {Bogdanov}, {Harrison}, {Hessels}, {Janssen},
  {Lyne}, {Patruno}, {Stappers}, {Stern}, {Tomsick}, {Boggs}, {Chakrabarty},
  {Christensen}, {Craig}, {Hailey}, \& {Zhang}}]{Tendulkar14}
{Tendulkar} S.~P. {et~al.}, 2014, ArXiv e-prints (astro-ph 1406.7009)

\bibitem[{{Wang} {et~al}\mbox{.}(2009){Wang}, {Archibald}, {Thorstensen},
  {Kaspi}, {Lorimer}, {Stairs}, \& {Ransom}}]{Wang09}
{Wang} Z., {Archibald} A.~M., {Thorstensen} J.~R., {Kaspi} V.~M., {Lorimer}
  D.~R., {Stairs} I., {Ransom} S.~M., 2009, \apj, 703, 2017

\bibitem[{{Wijnands}(2003)}]{Wijnands03c}
{Wijnands} R., 2003, \apj, 588, 425

\bibitem[{{Wijnands} {et~al}\mbox{.}(2001){Wijnands}, {Miller}, {Markwardt},
  {Lewin}, \& {van der Klis}}]{Wijnands01}
{Wijnands} R., {Miller} J.~M., {Markwardt} C., {Lewin} W.~H.~G., {van der Klis}
  M., 2001, \apjl, 560, L159

\end{thebibliography}

\end{document}